\documentclass{aa}  

\usepackage{graphicx}
\usepackage{txfonts}
\usepackage{subcaption}
\begin{document}

   \title{Binary origin of blue straggler stars in Galactic star clusters}

  % \subtitle{I. Overviewing the $\kappa$-mechanism}

   \author{ M.~J., Rain
          \inst{1}
          \and
          M.S., Pera\inst{3,4}
          \and
          G.I., Perren\inst{2,4}
          \and 
          O.G., Benvenuto\inst{2,5}
         \and 
         J.A., Panei \inst{2,5}
         \and 
         M.A. De Vito \inst{2,5}
         \and 
         G. Carraro \inst{6}
         \and
         S. Villanova\inst{7}
          }

   \institute{European Southern Observatory, Alonso de C\'ordova 3107, Vitacura, Región Metropolitana, Chile\\
              \email{mrainsep@eso.org}
         \and 
         Instituto de Astrof\'isica de La Plata, IALP (CONICET-UNLP), 1900 La Plata, Argentina
         \and 
         Instituto de F\'isica de Rosario, IFIR (CONICET-UNR), 2000 Rosario, Argentina
         \and 
         Facultad de Ciencias Exactas, Ingenier\'ia y Agrimensura (UNR), 2000 Rosario, Argentina
         \and 
            Facultad de Ciencias Astron\'omicas y Geof\'isicas, Universidad Nacional de La Plata (UNLP), La Plata, Argentina
        \and 
            Dipartimento di Fisica e Astronomia, Universit\`{a} di Padova, Vicolo Osservatorio 3, I-35122, Padova, Italy
        \and 
            Departamento de Astronom\'{\i}a, Universidad de Concepci\'{o}n, 160 Casilla, Concepci\'{o}n, Chile         
             }

   \date{Received September XX, XXXX; accepted March YY, YYYY}

% \abstract{}{}{}{}{} 
% 5 {} token are mandatory
 
  \abstract
   {Building on the recent release of a new \emph{Gaia}-based blue straggler star catalog in Galactic open star clusters (OCs), we explored the properties of these stars in a cluster sample spanning a wide range in fundamental parameters. We employed \emph{Gaia} EDR3 to assess the membership of any individual blue or yellow straggler to their parent cluster. We then made use of the \texttt{ASteCA} code to estimate the fundamental parameters of the selected clusters, in particular, the binary fraction. With all this at hand, we critically revisited the relation of the blue straggler population and the latter. For the first time, we found a correlation between the number of blue stragglers and the host cluster binary fraction and binaries. This supports the hypothesis that binary evolution is the most viable scenario of straggler formation in Galactic star clusters. The distribution of blue stragglers in the Gaia color-magnitude diagram was then compared with a suite of composite evolutionary sequences derived from binary evolutionary models that were run by exploring a range of binary parameters: age, mass ratio, period, and so forth. The excellent comparison between the bulk distribution of blue stragglers and the composite evolutionary sequences loci further supports the binary origin of most stragglers in OCs and paves the way for a detailed study of individual blue stragglers.}


   \maketitle
%
%-------------------------------------------------------------------

\section{Introduction}\label{sec:intro}

Defying the traditional single stellar evolution with their position in the optical color-magnitude diagram (CMD), they are bluer  and  brighter  than  the  main-sequence  turn-off  (MS-TO)  of  the  system  in  which  they  are  found, blue  straggler  stars  (BSSs) %\LEt{please check that you always add a plural s to "BSS" and "YSS" when required. I can see this sometimes, but not always}) 
and recently, yellow straggler stars (YSSs, which may be evolved BSSs) are  exotic  objects that have fascinated theorists and observers equally for generations. 
While  BSSs  were  initially  discovered  in  globular  clusters (GCs; \citealt{Piotto_2004, Salinas_2012}),  they  are  now  known  to  exist  in  open  clusters (OCs; \citealt{ Ahumada_2007, Rain_2021_2}),  dwarf galaxies \citep{Momany_2007},  and  even in  the  field of the Milky Way (e.g., \citealt{Santucci_2015}).  Since their detection on the core of globular cluster M3 \citep{Sandage_1953}, many formation  mechanisms  have  been  proposed. Most of them agree that a main sequence (MS) star has gained mass either through mass transfer (MT) from an evolving primary star via Roche-lobe overflow (RLOF) \citep{McCrea_1964} and/or via collisions involving single, binary, or even triple stars \citep{Hills_1976}. The two basic scenarios can be modified by the presence of a third or additional star. \cite{Perets_Fabrycky_2009} proposed a scenario in which the inner binary in a hierarchical triple system can coalesce due to perturbations of the outer companion, producing in the end a BSS in a binary system with a long period.  \\

With the advent of the second \emph{Gaia} data release, a renaissance of the study of BSSs on individuals and across a large sample of OCs took recently place \citep{Bhattacharya_2019, Rain_2020, Vaidya_2020, Rain_2021_1, Jadhav_2021, Leiner_2021, Rain_2021_2, Jadhav_2021, Rao_2023, Rani_2023}. It is possible today to accurately identify genuine BSSs candidates while distinguishing them from outliers and field stars by combining \emph{Gaia} parallax measurements, proper motions, and star colors to establish membership with a high degree of certainty. 
In this study, we use \emph{Gaia} EDR3 to select members of a sample of 12 old ($>$ 9.0 Gyr), relatively nearby (d$<$ 5000 parsecs), rich OCs, with the goal of understanding the %\LEt{you introduced and used an abbreviation for this above. Please decide whether you wish to spell out or abbreviate and then change throughout accordingly for consistency. This applies to all abbreviations and acronyms: they are to be introduced at first occurrence in the main text and are then to be used without reintroduction for consistency, including all figure captions and tables. Please check this throughout and change as required. I'll not highlight this again to avoid cluttering the ms.} 
BS population and their connection with binaries in the host cluster. 

The data on BSSs and YSSs allow us to perform a detailed comparison with theoretical predictions. For this purpose, we present detailed calculations of binary evolution to explore whether the hypothesis of a binary origin holds. The quantities necessary to define a particular binary are the stellar masses, the orbital period of the pair, and the type of mass transfer, conservative or nonconservative. Thus, performing a full exploration of the parameter space of the problem represents a major numerical effort that may be warranted in a future study. We restrict ourselves here to a given (fixed) initial mass ratio and conservative MT.

The layout of this study is as follows: We first describe our selection of the sample and the cluster members in \S~\ref{sec:data_selection}, and in \S~\ref{section:bs_ys_region}, we define the regions whithin which the blue and yellow straggler candidates where selected. In \S~\ref{sec:asteca} we describe the estimation of fundamental parameters. Then, in \S~\ref{sec:correlations}, we search for correlations between the cluster parameters and the straggler population.  In \S~\ref{sec:Binary_Evolution} we introduce binary evolution models, and in \S~\ref{sec:comparison} we compare them with the blue straggler star distribution in the color-magnitude diagram. In \S~\ref{sec:conclusions} we provide our summary and conclusions.

\begin{figure}
\centering
\includegraphics[width=\linewidth]{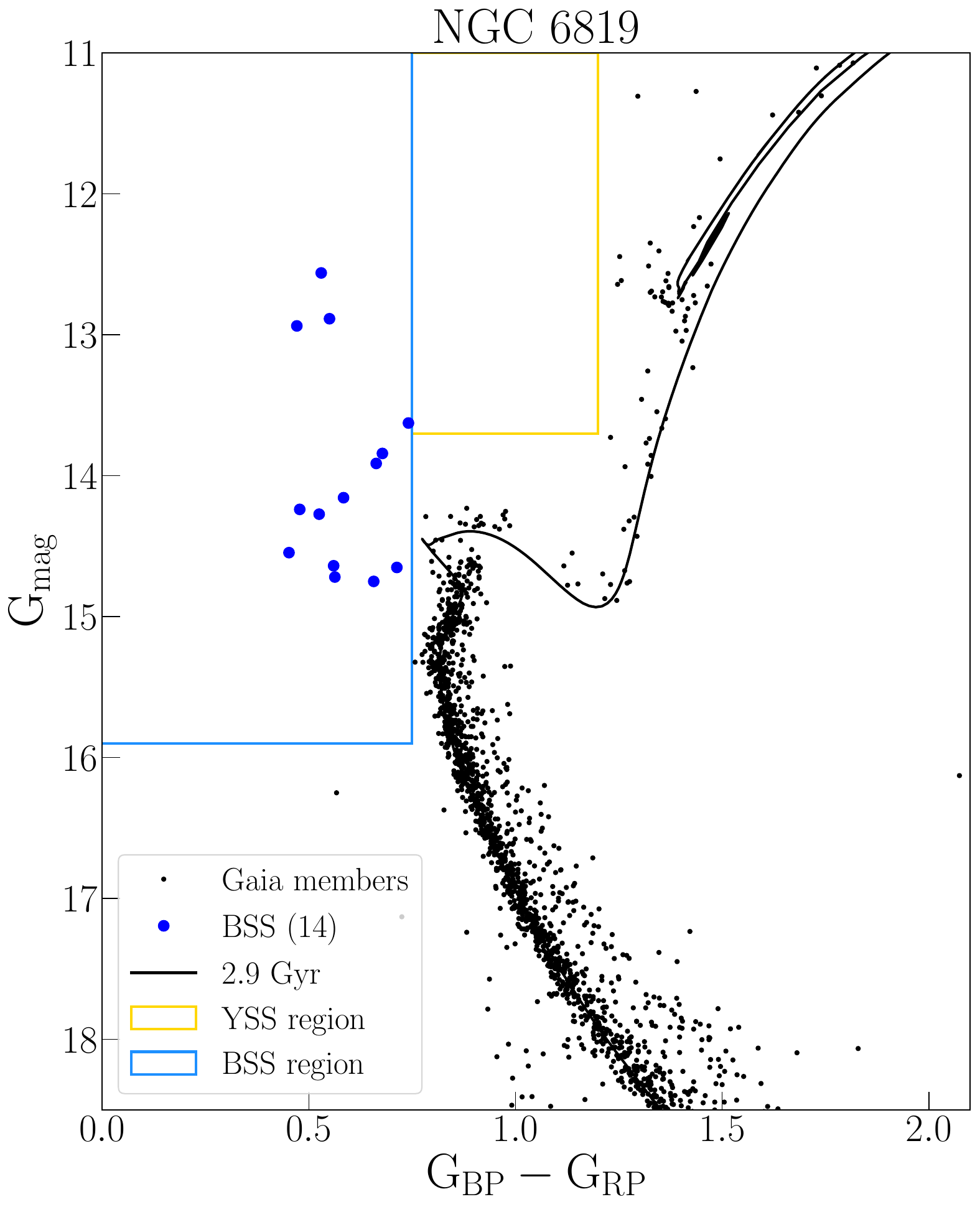}
\caption{Color-magnitude diagram of the open cluster NGC~6819 to illustrate the selection of BSSs and YSSs. The light blue and yellow lines separate the different regions in which we searched for these stars. The MIST isochrone is shown for comparison. }
\label{fig:cmd_example}
\end{figure}

\begin{table*}
\centering
\caption{Parameters of open clusters in our sample adopted for this study.} \label{tab:general}
\begin{tabular}{lccccc}
 \hline \hline
Cluster & log(age) & $\mathrm{f_{bin}}$ & Mass & (m-M)$_{0}$\\
 & & & [M$_{\odot}$] & & \\
\hline
 Berkeley~32      & 9.84$\pm$0.00  & 0.29$\pm$0.01 & 16649$\pm$559  & 12.52$\pm$0.00\\
 Berkeley~39      & 9.89$\pm$0.02  & 0.23$\pm$0.04 & 16716$\pm$967  & 13.10$\pm$0.02\\
 Collinder~261    & 9.90$\pm$0.03  & 0.37$\pm$0.02 & 36941$\pm$2600 & 12.10$\pm$0.01\\
 Melotte~66       & 9.67$\pm$0.00  & 0.34$\pm$0.02 & 20958$\pm$887  & 13.20$\pm$0.00\\
 NGC~188          & 9.89$\pm$0.00  & 0.32$\pm$0.03 & 14860$\pm$607  & 11.20$\pm$0.00\\
 NGC~2141         & 9.41$\pm$0.00  & 0.36$\pm$0.02 & 26251$\pm$1385 & 13.10$\pm$0.01\\
 NGC~2158         & 9.48$\pm$0.12  & 0.36$\pm$0.05 & 38854$\pm$4453 & 12.77$\pm$0.06\\
 NGC~2243         & 9.54$\pm$0.00  & 0.15$\pm$0.02 & 9279$\pm$888   & 13.00$\pm$0.00\\
 NGC~2506         & 9.32$\pm$0.02  & 0.19$\pm$0.02 & 19934$\pm$1194 & 12.49$\pm$0.02\\
 NGC~2682         & 9.71$\pm$0.02  & 0.23$\pm$0.04 & 7437$\pm$533   & 9.20$\pm$0.00\\
 NGC~6819         & 9.66$\pm$0.06  & 0.17$\pm$0.02 & 28157$\pm$1326 & 11.80$\pm$0.03\\
 NGC~7789         & 9.20$\pm$0.00  & 0.24$\pm$0.02 & 22150$\pm$778  & 11.20$\pm$0.00\\
\hline \hline
\end{tabular}
\end{table*}

\section{Selection of the cluster sample and members}\label{sec:data_selection}

The clusters were first selected on the basis of the number of blue stragglers. Only those with N$_{BSS} \geq 8$ listed in the most recently published catalog of BSSs in OCs  \citep{Rain_2021_2} were selected. The original list contained a total of 32 clusters with $\log(\mathrm{age})\geq 9.0$~dex~($1~\mathrm{Gyr}$, \citealt{Dias_2021}), distances d $>$ 850 pc~\citep{Cantat-Gaudin_2020}, and masses M $> 1400~\mathrm{M_{\odot}}$~\citep{Jadhav_2021}.

For the 32 clusters, we performed a membership selection using the Gaia astrometric solution. The cluster members were estimated via a two-step process.
First, we downloaded the data of each cluster by using a simple script to query EDR3 data using the \texttt{Astroquery} package\footnote{https://github.com/Gabriel-p/GaiaQuery}. This package generates two user-defined colors (not present in the raw Gaia data) with their associated uncertainties. This is currently not provided in \emph{Gaia}. The last was useful for the process described in Section 4 (\S~\ref{sec:asteca}). For each cluster, we then selected all the stars within twice the apparent radius reported in \cite{Dias_2002} and with magnitudes down to $\mathrm{G_{mag}}=18.5$.
Second, the most likely members for each cluster were identified with the \texttt{pyUPMASK} code~\citep{Pera_2021}. This tool assigns probability membership $\mathrm{P_{memb}}$ values to all the stars in an observed frame based on input data selected by the user.  In this case, we employed parallax and proper motion data that we downloaded from the \emph{Gaia}~EDR3 survey as the input data.\\
After assigning probabilities, the final member list was automatically generated by an iterative process. This process works by filtering out low-probability stars at each step and halts when the density of stars within the cluster region is consistent with the expected density, taking into account the field density value outside the adopted cluster region. 
After this membership analysis, we removed nine clusters from our list because of their sparse nature. With only a handful of subgiant (SGB) and read giant branch (RGB) stars, these clusters have large errors for the few observed stars and can lead to an overestimation of the parameters (e.g., mass and age), and therefore, we did not consider them reliable enough for the main aim of this work.

%\subsection{Reddening corrections}\label{subsec:diff_reddening}

In particular, the \emph{Gaia} photometric bands are broad enough to introduce large color differences  caused by extinction as a function of the stellar SED. These spreads in color can introduce dispersion in the CMD positions, affecting the selection, especially  near the  TO.  As shown by \cite{Leiner_2021}, for clusters with low reddening values ($E(G_\mathrm{BP}-G_\mathrm{RP}) < 0.3$), it is sufficient to  adopt the reddening from the literature and convert it into the \emph{Gaia} passbands. On the other hand, for clusters with higher values of $E(G_\mathrm{BP}-G_\mathrm{RP})$, individual (star by star) reddening corrections are recommended.

In \cite{Rain_2021_1} we determined the reddening law $R_{G}=A_{G}/E(G_\mathrm{BP}-G_\mathrm{RP})$ by a linear least-squares fit and obtained $R_{G}=1.79\pm 0.05$. This value was adopted as the slope of the reddening law for all the clusters. Here, we carried out the reddening corrections using MS stars. For these, we defined a line along the MS, and for each one of the selected MS stars, we calculated its distance from this line to the reddening law line. The vertical projection of this distance gives the differential $A_{G}$ absorption at the position of the star, while the horizontal projection gives the differential reddening $E(G_\mathrm{BP}-G_\mathrm{RP})$ at the stellar position. After this first step, we selected for each star of the field (both cluster members and nonmembers) the ten nearest MS stars and calculated the mean differential $A_{G}$ absorption and the mean differential $E(G_\mathrm{BP}-G_\mathrm{RP})$, and we finally subtracted this mean value from its  ($G_\mathrm{BP}-G_\mathrm{RP}$) color and G magnitude. Although they are sufficiently rich, seven clusters situated beyond 5.5 Kpc and with an extinction between $1.0 < \mathrm{A_{v}} < 2.5$ exhibit a considerable variation in the color of their MS and giant branches, particularly in the TO region. As a result, it becomes very challenging to accurately fit isochrones and define the blue (and yellow) region, as explained in \S~\ref{section:bs_ys_region}. 

\section{Blue and yellow straggler regions}\label{section:bs_ys_region}
Although a great variety of definitions is available in the literature \citep[and references therein]{Ahumada_2007, Bhattacharya_2019, Rain_2020, Vaidya_2020} to define the blue straggler region, we closely followed the most recent definitions using \emph{Gaia} data \citep{ Leiner_2021, Rain_2021_2}. The procedure we used is the following: i) After the membership selection (see Section \S~\ref{sec:data_selection} ), the photometric data were plotted in the  $G$ versus ($G_\mathrm{BP}-G_\mathrm{RP}$)  diagram. ii) An approximate matching of a \textsc{MIST} theoretical isochrone  \citep{Dotter_2016} with the \emph{Gaia}~EDR3 passbands and assuming the parameters of each cluster calculated by \cite{Dias_2021} was then performed on the main sequence and TO, and eventually on the RGB and red clump (RC), if present. iii) Then, we identified the bluest point in the best-fit isochrone. We used the blue hook when available, and otherwise, the TO to set this limit. Only objects significantly detached from this point, for instance, $\sim$ 0.03 mag as a minimum and down to 0.5 mag at most below the TO, were listed as blue straggler candidates. iv) We plotted the equal-mass binary sequence obtained by displacing the isochrone by 0.75~mag upward, which represents the maximum brightness expected for an equal-mass binary at the cluster TO location. The very same sequence was helpful to identify the yellow straggler stars and also to set the lower limit (on brightness) of this population.  Stars brighter than this sequence but redder than the blue stragglers and bluer than the red giant branch falling into this region were considered yellow straggler candidates (see Figure \ref{fig:cmd_example}).  \emph{Gaia} CMDs of the remaining clusters are shown in Appendix \ref{append_sec:cmds}. After this step, only clusters with $\mathrm{N_{BSS}} \geq 9$ remained in our list. Four clusters were left out because they had a relatively low number of stragglers.

\section{Estimation of the fundamental parameters with ASteCA}
\label{sec:asteca}

The fundamental parameters for all the clusters (age, binary fraction, distance,
metallicity, extinction, and mass) were estimated  with the
\texttt{ASteCA} package~\citep{Perren_2015}. This package has been used to
successfully analyze hundreds of clusters since its
release~\citep{Perren_2017,Perren_2020}. To simplify the Bayesian inference,
the process that estimates the fundamental parameters, we assumed a general solar
metallicity and allowed the rest of the parameters to vary within reasonable
ranges. The estimation of the binary fraction depends on the chosen distribution
of primary to secondary masses $q=m_1/m_2$ (mass ratio), where $m_1$ is the
primary mass of a binary system, and $m_2$ is the secondary mass.
Here, the distribution is specified to be uniform as

\begin{equation}
  q=\begin{cases}
    0, & \text{$q>q_\mathrm{max}$}\\
    1, & \text{$q\leq q_\mathrm{max}$}
  \end{cases}
\end{equation}
with $q_\mathrm{max}= 1.43$. The shape of these distributions approximates the empirical distributions found in works such as \cite{Fisher_2005} and \cite{Raghavan_2010}, where the mass ratio is close to unity ($m_{2}\approx m_{1}$) and drops rapidly for lower values ($m_{2}\le 0.5
1m_{1}$). The results for each cluster are reported in Table \ref{tab:general}.
At this point. it is important to mention that opting for a uniform distribution would lead to a higher binary fraction value because of the increased production of binary systems with lower secondary masses. Hence, it is appropriate to view our binary fraction value as a conservative lower limit.

\begin{figure}
\centering
\includegraphics[width=\linewidth]{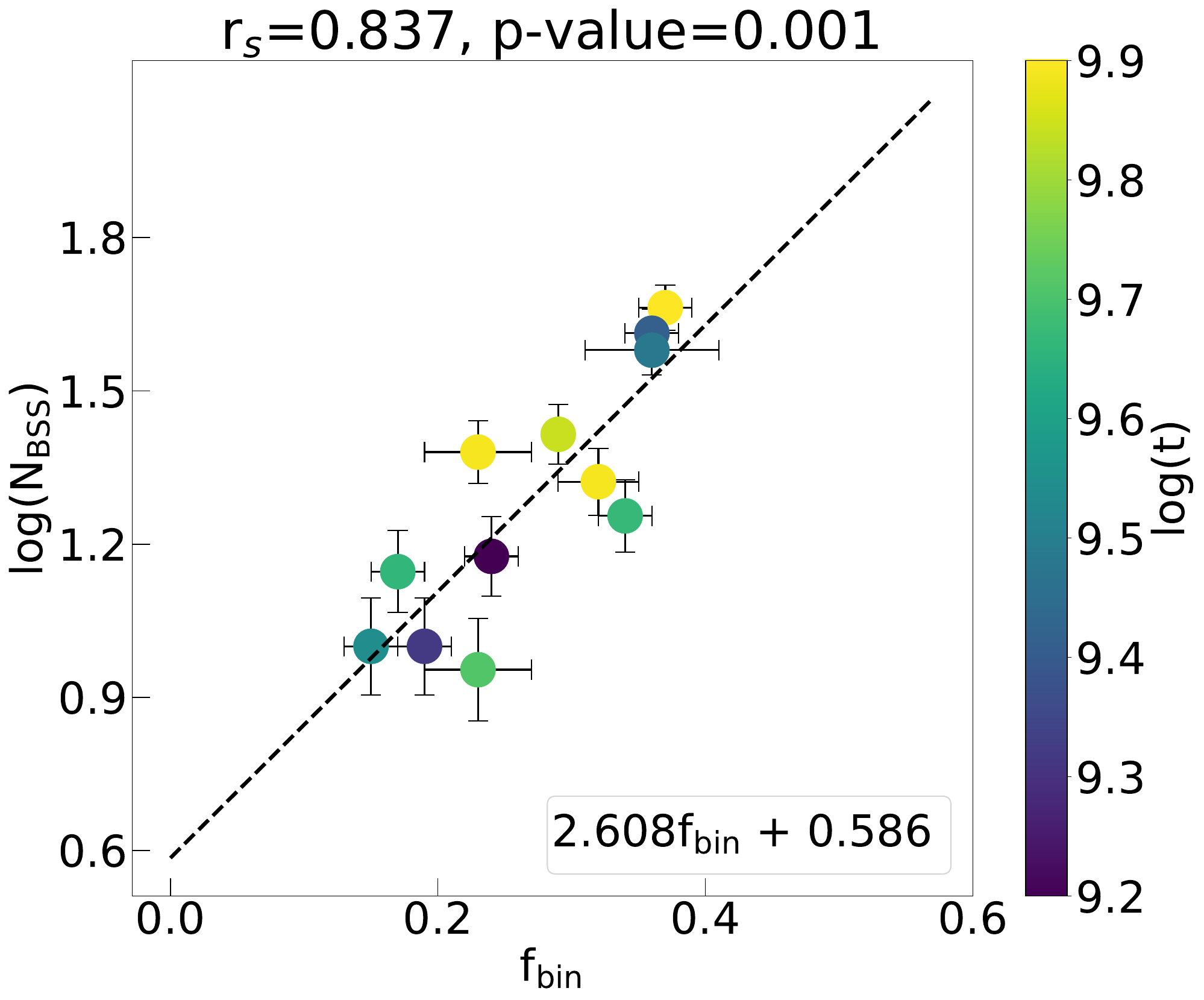}
\caption{Logarithm of the observed number of blue straggler stars a function of the binary fraction. The dashed black line indicates the best fit to the data. The color bar indicates the age of each cluster. Errors are Poissonian.}\label{fig:fb_nbs-nx}
\end{figure}

\section{Searching for correlations}\label{sec:correlations}

\begin{figure}
\centering
\includegraphics[width=\linewidth]{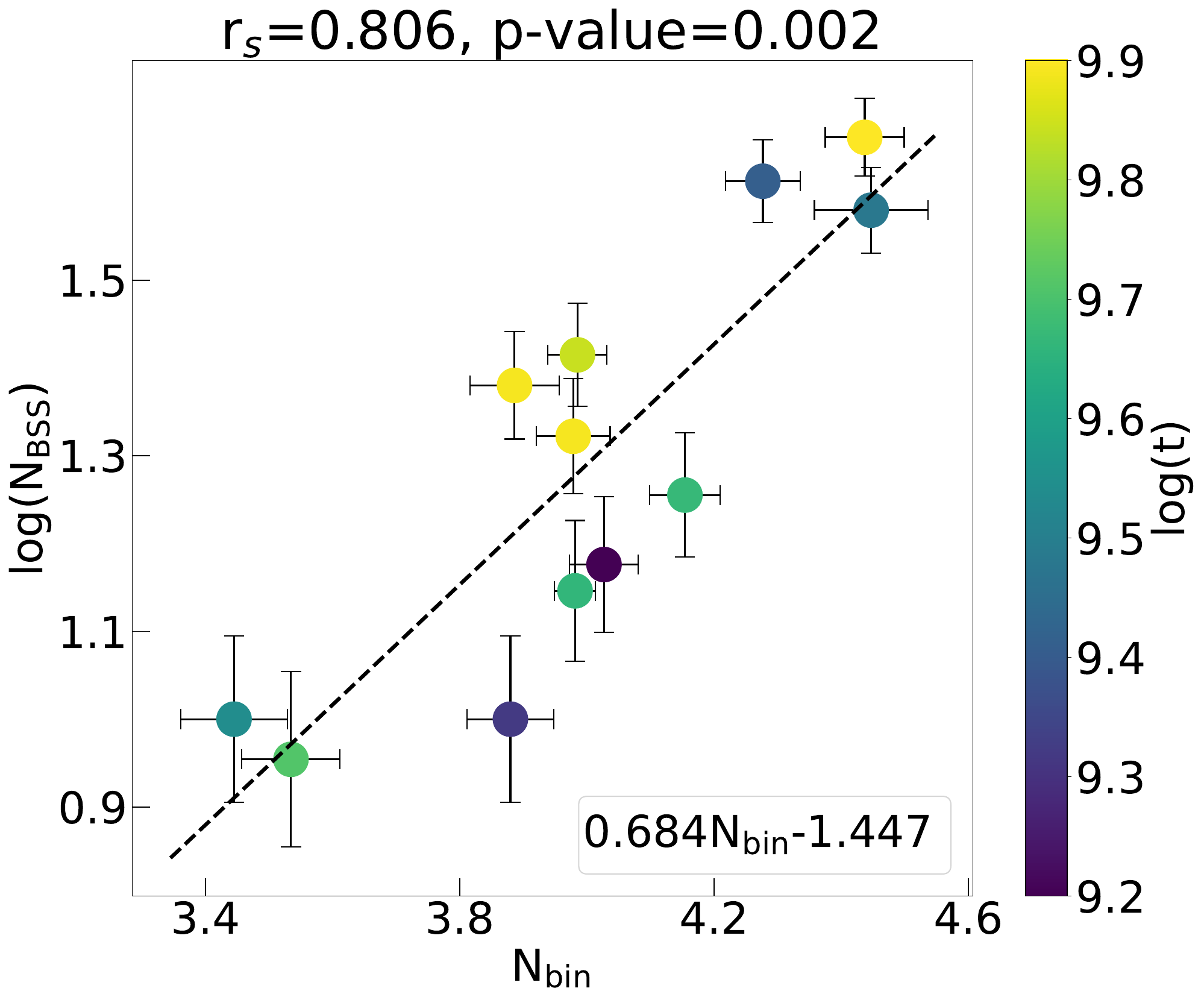}
\caption{Logarithm of the observed number of BSSs as a function of the number of binaries on each cluster. The dashed black line indicates the best fit to the data. The errors are Poissonian.}
\label{fig:Nbin_Nbs}
\end{figure}

An alternative way of obtaining insight into the relative efficiency of the formation mechanisms of BSSs is to investigate possible correlations between the blue straggler population and the host cluster properties. We therefore compared the observed number of BSSs ($\mathrm{N_{BSS}}$) with the physical parameters of the 12 OCs of the sample. In particular, we searched for possible correlations between the raw number of BSSs (N$_\mathrm{{BSS}}$), %age in logarithm scale $\log(\mathrm{age})$, 
the binary fraction ($\mathrm{f_{bin}}$), and the number of binaries $\mathrm{N_{bin}}$ of each cluster. Furthermore, to statistically measure the strength of the relation between the variables, we used the Spearman rank correlation test. The coefficient ($r_{s}$) and confidence levels (p-value) for the considered parameter pairs are reported in Figure \ref{fig:fb_nbs-nx} and Figure \ref{fig:Nbin_Nbs}.

We revisited the correlation between $\mathrm{N_{BSS}}$ and cluster binary fraction. Open clusters are both less massive and younger than globular clusters. Their proximity and low density make the determination of the binary frequency particularly easy. Furthermore, in open clusters, the relation between this parameter and straggler numbers has never been explored in detail before. 
We find a dependence $\mathrm{N_{BSS}} \propto \mathrm{f_{bin}}^{0.5\pm0.11}$ and a Spearman coefficient value of $r_{s}$=0.83. This correlation is shown in Figure \ref{fig:fb_nbs-nx} with the corresponding line of the best fit to the data.  As in the case of the GCs, here, the relation is continuous \citep{Milone_2012} and can be considered as the continuation of that found between core binary fraction and straggler numbers in low-density clusters by \cite{Sollima_2008}.

Within the same context, the cluster mass inside the core radius has been considered by different authors to be the best indicator of blue straggler population size in stellar clusters (e.g., \citealt{Knigge_2009,Leigh_sills_2011}).  We found a dependence of $\mathrm{N_{BSS}} \propto \mathrm{M_{Tot}^{\delta}}$ on $ \delta$=0.6$\pm$0.2 here. These values are higher than the predicted value reported on \cite{Knigge_2009} ($\mathrm{M_{tot}^{\delta=0.4-0.5}}$) for GCs, but they agree with the recent upper limit found for OCs $\mathrm{M_{Tot}}^{\delta=0.6}$ \citep{Jadhav_2021}. 

According to \cite{Leigh_2013}, when most of the BSSs are descended from binary evolution, a dependence of the form
\begin{equation}\label{eq_2}
\mathrm{N_{BSS} \propto \mathrm{N_{bin}} \sim \frac{\mathrm{f_{bin}} M_{tot}}{\overline{m}}} 
\end{equation}
is expected, where $\overline{m}$ is the average stellar mass, which the authors assumed to be the same for every cluster. We assumed this to be 0.4~M$_{\sun}$.

In comparison with the binary fraction, the correlation with the N$_{\mathrm{bin}}$ decreases. When using Equation~(\ref{eq_2}), the Spearman rank coefficient $\mathrm{r_{s}}$ decreased in $\sim$ 0.03. This means that the correlation becomes weaker by adding $\mathrm{M_{Tot}}$ (see Figure~\ref{fig:Nbin_Nbs}). This behavior was predicted by \cite{Knigge_2009} and previously tested by \cite{Leigh_2013} in GCs. Our results agree with their findings, that is, the strength of the correlation ($\mathrm{N_{BSS}}$ vs. $\mathrm{M_{core}}$) decreases when $\mathrm{f_{bin}}$ is included. Within our sample, we found a dependence of the form $\mathrm{N_{BSS}} \propto \mathrm{N_{bin}^{0.75\pm0.13}}$.

Finally, and differently from the previous cases, no correlation at all was found between the specific frequency ($\mathrm{F}_{\mathrm{MS}}^{\mathrm{BSS}}$) \footnote{$\mathrm{F}_{\mathrm{MS}}^{\mathrm{BSS}}$ = log(N$_\mathrm{{BSS}}$/N$_\mathrm{{MS}})$} and $\mathrm{M_{core}}$ or N$_{\mathrm{bin}}$. The absence of a correlation between these parameters is expected, however, because according to \cite{Knigge_2009}, Equation~(\ref{eq_2}) only holds for the blue straggler numbers and not for the specific frequency.

\begin{figure*}
\centering
\includegraphics[width=250pt]{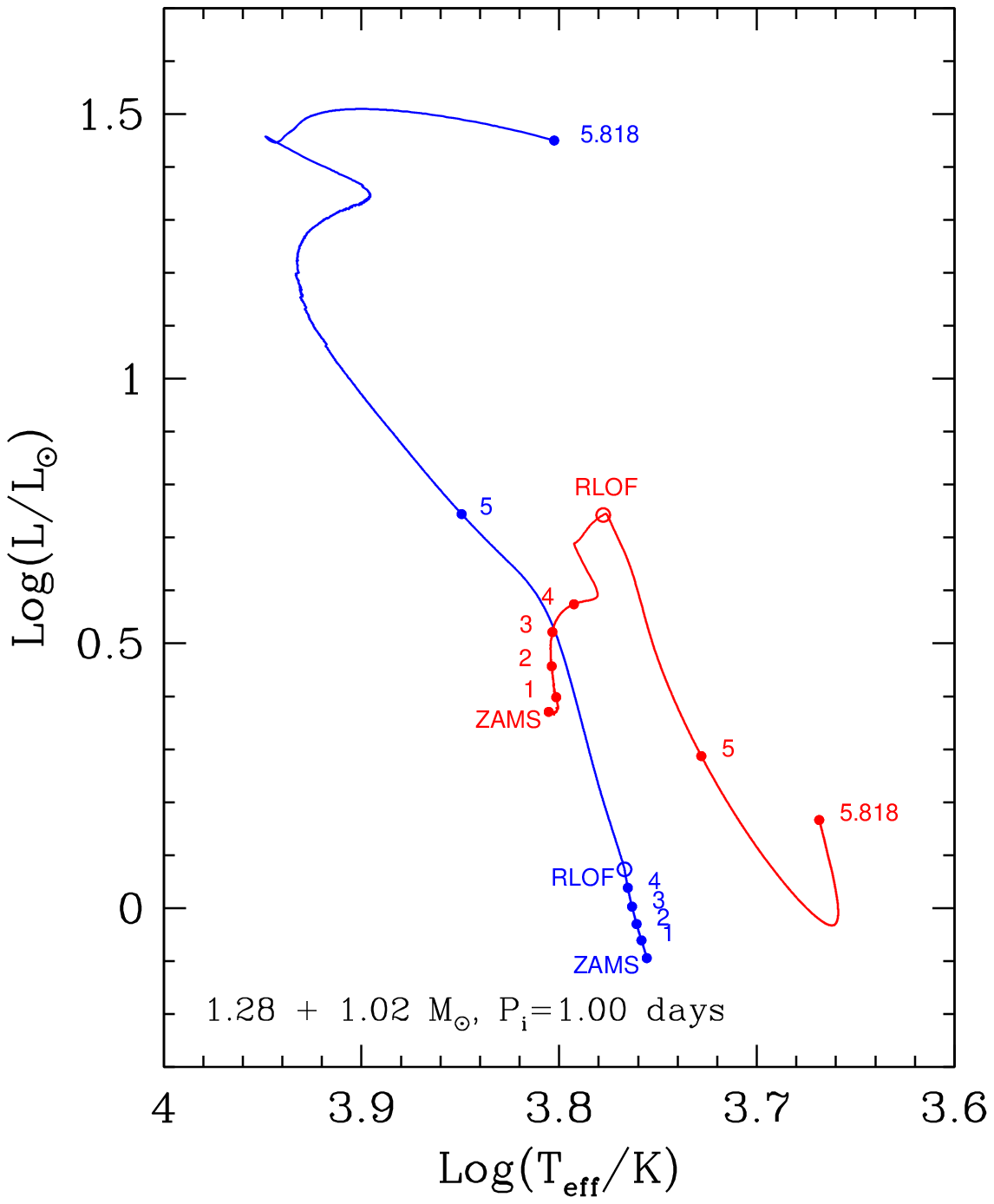}
\includegraphics[width=250pt]{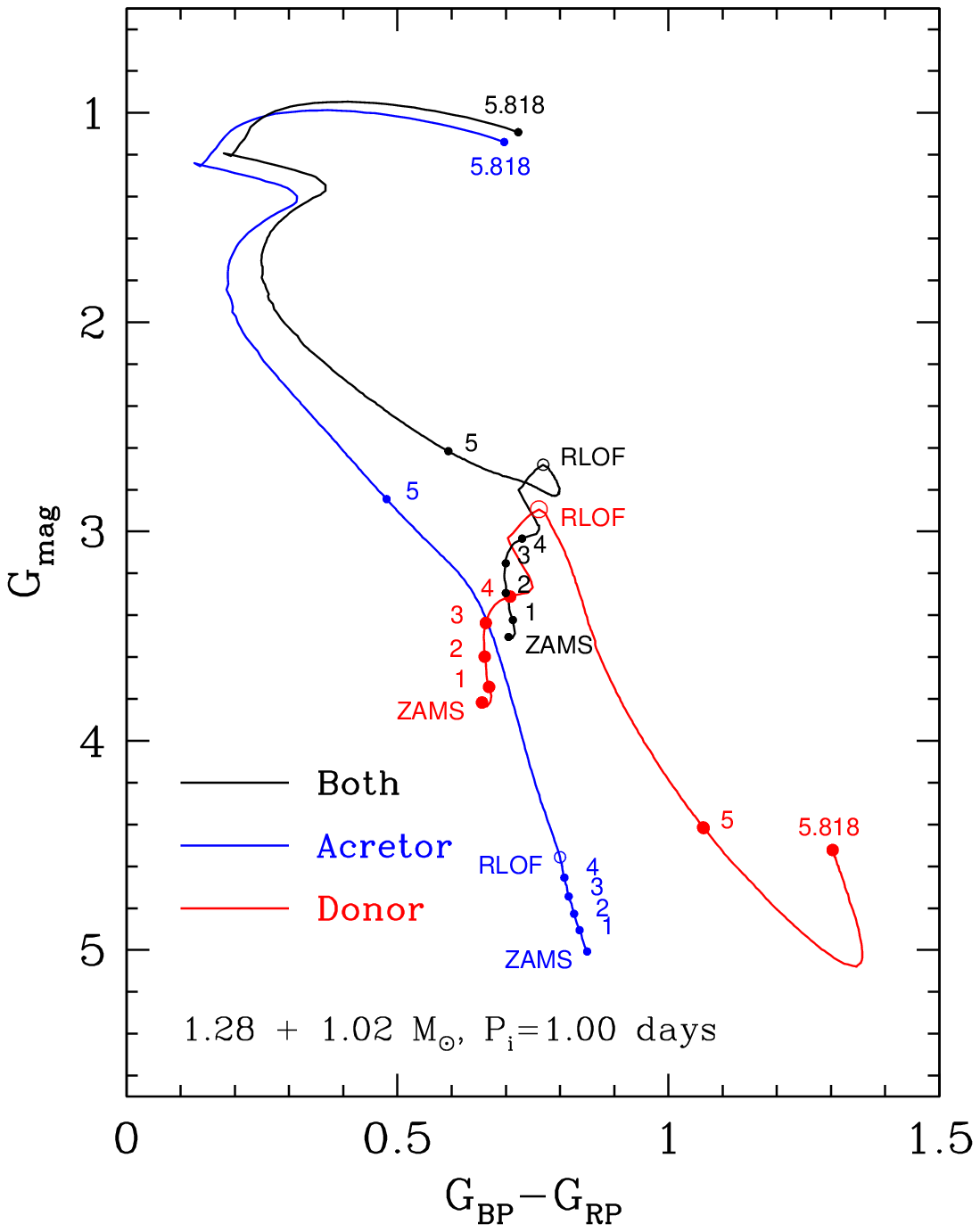}
\caption{Evolution of a pair of 1.28~$M_{\odot}$~+~1.02~$M_{\odot}$ on an orbit with an initial period of one day. The filled points are labeled with their corresponding ages. The open circles denote the occurrence of the RLOF at an age of $\approx$~4.9~Gyr. At this moment, it has already exhausted its hydrogen core. \textbf{Left panel}: Evolution of the components of the pair in the theoretical plane of bolometric luminosity vs. effective temperature. \textbf{Right panel}: Each track in the CMD and also the track to be observed by a remote observer that cannot resolve the pair.}
\label{fig:Evolu_example}
\end{figure*}

\section{Binary evolution and the formation of blue straggler stars} \label{sec:Binary_Evolution}

To theoretically account for the observed properties of BSSs,  we performed a set of detailed binary evolution calculations. The results presented  below represent our first attempt to reproduce them.

When considering binary evolution calculations, a set of quantities has to be defined (the initial masses of the stars and the orbital period; the type of mass transfer, see below, and the chemical composition, etc.). Thus, performing a full exploration of the parameter space represents a major numerical effort.

For the calculations presented in this paper, we employed an updated version of the code for stellar binary evolution described in \citet{2003MNRAS.342...50B}. Briefly, this code solves the structure of spherical stars assuming they move along a circular orbit. It allows the computation of conservative and nonconservative mass transfer episodes. When the pair is detached, it works as a standard Henyey code. Conversely, hen the donor has a size comparable to that of its Roche lobe and a mass transfer sets in (a semidetached pair), it computes the structure of the donor, the orbital evolution, and the mass transfer rate simultaneously. %\typeout{OGB: No se, creo que con lo que esta escrito es suficiente \ale{Deberiamos aclarar que la tasa de transferencia de masa entra como una variable en el cálculo evolutivo??? Esta descripto en el trabajo citado, pero puede que valga la pena recalcarlo.}}

The calculations were performed assuming that there is no mass loss from the system: All the material transferred by the donor is accreted by its companion. This process causes the system to become a blue straggler and eventually, a yellow straggler. Conservation of mass represents an extreme case of binary evolution in which the BSSs phenomenon is strongest. Evidently, if the companions were able to retain only a fraction of the transferred mass, the increase in mass and luminosity would be smaller.

In order to compute the evolution of a binary system, we first considered the evolution of the donor star. By doing this, we determined the structure of this star, the mass transfer rate, the chemical composition of the material lost by the donor, and the orbital evolution. After this, with the mass transfer rate ($\dot{M}$) from the previous calculation, we performed the detailed evolution of the companion star (the accretor). As a result of accretion and internal evolution, this star filled its own Roche lobe in most cases. At this moment, the components of the system were in contact, and we stopped the calculations. The object that emerged from contact is expected to be one of those observed in any case.

We tuned the mixing length parameter to reproduce the present Sun, considered moderate overshooting, semiconvection as in \citet{1983A&A...126..207L}, and thermohaline mixing following \citet{2013A&A...553A...1M}. The chemical evolution of the models was solved as in \citet{1985A&A...145..179L} considering noninstantaneous mixing. We neglected effects of stellar rotation in these calculations.

We assumed solar composition stars. The donor masses ranged from 0.82~$M_{\odot}$ to 1.60~$M_{\odot}$ with a step of 25\%; for the accretor, we simply assumed an initial mass ratio of 1.25, and the initial orbital period interval extended from 0.26~d to 1.95~d, again with a step of 25\% (see Table~\ref{tab:binary_models}). 

The choice of the initial mass ratio of 1.25 was made to perform the first step of our exploration of the BSSs phenomenon. We designed these calculations with the aim of quantitatively testing the correctness of the hypotheses we developed to explain BSSs in open clusters. A more exhaustive exploration of possible configurations with other initial mass ratios will be the subject of future investigations.

Stellar evolution calculations provide the bolometric luminosity and effective temperature of each star of the pair, among other quantities. However, observations do not detect the donor and the accretor separately, but together as one object. To allow a direct comparison of models with data, after solving for the evolution of each star of the pair, we therefore added their contributions to compute its evolution in the theoretical CMD (see below \S\ref{subsec:colores}). These results are to be compared with observations.

\subsection{Evolutionary results} \label{subsec:evol_results}

Fig.~\ref{fig:Evolu_example} shows an example of our results. We show the evolution of a pair of 1.28~$M_{\odot}$~+~1.02~$M_{\odot}$ on an orbit with an initial period of one day. The system evolves on a detached configuration up to an age of $\approx$~4.9~Gyr, when the donor star fills its Roche lobe. At this moment, it has already exhausted its hydrogen core. It accordingly undergoes class~B mass transfer. From then on, the star follows a completely different evolution from that of an isolated star of the same mass and composition. Prior to RLOF, the companion star evolves very slowly, but when it begins to accrete mass, it becomes brighter on a very short (thermal) timescale. The evolution was followed up to an age of 5.89~Gyr when the accretor also fills its Roche lobe and the system is in contact.

In the left panel of Fig.~\ref{fig:Evolu_example}, we depict the evolution of the system in the typical theoretical plane, and in the right panel, we show each of the tracks in the CMD (see \S~\ref{subsec:colores} below). In this panel, we also include the track to be observed by a remote observer that cannot resolve the pair. We call this sequence a composite evolutionary sequence.

In Table~\ref{tab:binary_models} we list the computed binaries, where we indicate with $M_{A}$ and $M_{B}$ the masses of donor and accretor stars in solar mass units, respectively, and the initial orbital period in days.

In closing this section, a comparison of our work with approaches and results presented by other authors is in order.\citet{Tian_2006} studied the formation of BSSs by mass transfer in close binary systems in a somewhat similar way to the way adopted in this paper. Employing the Eggleton code  \citep{2000MNRAS.319..215H} and assuming conservative mass transfer, they computed a set of binary systems for a variety of donor and accretor masses and orbital separations.  They applied this set to perform a population study for the conditions of the old OC NGC~2682 (M67), concluding that at least for the case this OC, other mechanisms must operate that lead to BSSs formation in addition to mass transfer in binaries. Among other more recent works, we cite those of \citet{2021ApJ...908....7S} and \citet{2023ApJ...944...89S}. In each of these papers, the authors presented a study of a particular BSS (WOCS 5379 and WOCS 4540, both located in the OC NGC~188) based on detailed models constructed with the MESA code \citep{2011ApJS..192....3P}, searching for the models that best account for their main characteristics. They allowed for the possibility of nonconservative mass transfer. Another study, presented by \citet{Leiner_2021}, considered a population analysis of BSSs based on simulations performed with the rapid code presented by \citet{2002MNRAS.329..897H}, which allows approximately computing a large number of systems.  Binary stellar evolution strongly depends on more parameters (masses, orbital period, the type of mass transfer, and chemical composition) than isolated objects, but also on the stability of the mass transfer process transfer. The results presented by \citet{Leiner_2021} strongly depend on some approximations that strongly affect the output of the calculations, in particular, the critical mass ratio for stable mass transfer $q_{crit}$.

We compared our effort with those presented in the four papers specifically devoted to BSSs that we cited above. In this paper, we do not try to fit models to a particular BSS as in \citet{2021ApJ...908....7S} or \citet{2023ApJ...944...89S}, but search for a wider view of the problem. While our models are detailed, they are unable to provide a picture of the problem as wide as that given by the population analysis of \citet{Leiner_2021}. The approach assumed by \citet{Tian_2006} is more similar to ours, although there are some remarkable differences. They computed a family of detailed evolutionary tracks under similar assumptions regarding the mass transfer process (conservation of mass and angular momentum) for a variety of masses and initial periods and employed them in a population study for NGC~2682. We worked with a fixed pair-mass, however, which allowed a variety of initial periods, and we applied it to a variety of clusters, but did not simulate any population synthesis. 

The detailed comparison of our models with those presented in the above-cited works is an important and not easy task. In order to avoid deviating from the main objective of this paper, we defer this analysis to a future publication.

\begin{table}
%\tablenum{4}
\centering
\caption{Binary models. List of binaries computed: Donor mass (M$_A$), accretor mass (M$_B$), and initial orbital period.} \label{tab:binary_models}
\begin{tabular}{ccc|cccc}
\hline \hline
%\tablewidth{0pt}
$M_{A}$ & $M_{B}$ & Period & & $M_{A}$  & $M_{B}$ & Period\\
M$_{\odot}$ &M$_{\odot}$ & days & & M$_{\odot}$ & M$_{\odot}$ & days \\
\hline

%\decimalcolnumbers
0.82   &  0.65   &  0.26  & &  1.28  &  1.02   &  0.76 \\
 "     &  "      &  0.27  & &  "     &  "      &  0.80 \\
 "     &  "      &  0.29  & &  "     &  "      &  0.85 \\
1.02   &  0.81   &  0.51  & &  "     &  "      &  0.90 \\
 "     &  "      &  0.58  & &  "     &  "      &  0.95 \\
 "     &  "      &  0.64  & &  "     &  "      &  1.00 \\
 "     &  "      &  0.68  & &  1.60  &  1.28   &  0.51 \\
 "     &  "      &  0.72  & &  "     &  "      &  0.64 \\
 "     &  "      &  0.76  & &  "     &  "      &  0.80 \\
 "     &  "      &  0.80  & &  "     &  "      &  1.00 \\
1.08   &  0.87   &  0.51  & &  "     &  "      &  1.13 \\
1.15   &  0.92   &  0.51  & &  "     &  "      &  1.19 \\
 "     &  "      &  0.64  & &  "     &  "      &  1.25 \\
 "     &  "      &  0.80  & &  "     &  "      &  1.33 \\
1.28   &  1.02   &  0.51  & &  "     &  "      &  1.41 \\
 "     &  "      &  0.58  & &  "     &  "      &  1.48 \\
 "     &  "      &  0.64  & &  "     &  "      &  1.56 \\
 "     &  "      &  0.68  & &  "     &  "      &  1.95 \\
 "     &  "      &  0.72  & &        &         &       \\
\end{tabular}
\end{table}

\subsection{Colors and magnitudes in the Gaia system} \label{subsec:colores}

For all our binaries, we computed Gaia EDR3 filters to perform
color-magnitude diagrams. To this end, we solved the atmosphere structure of our models in local thermodynamic equilibrium. The code we used is an updated version of the code described in \citet{Rohrmann2001MNRAS.323..699R}. It takes into account hydrostatic and radiative-convective equilibrium. When the star model had a cold atmosphere, convective transport was treated with the mixing-length approximation. 

\begin{figure}
\centering
\includegraphics[angle=270, width=0.45\textwidth]{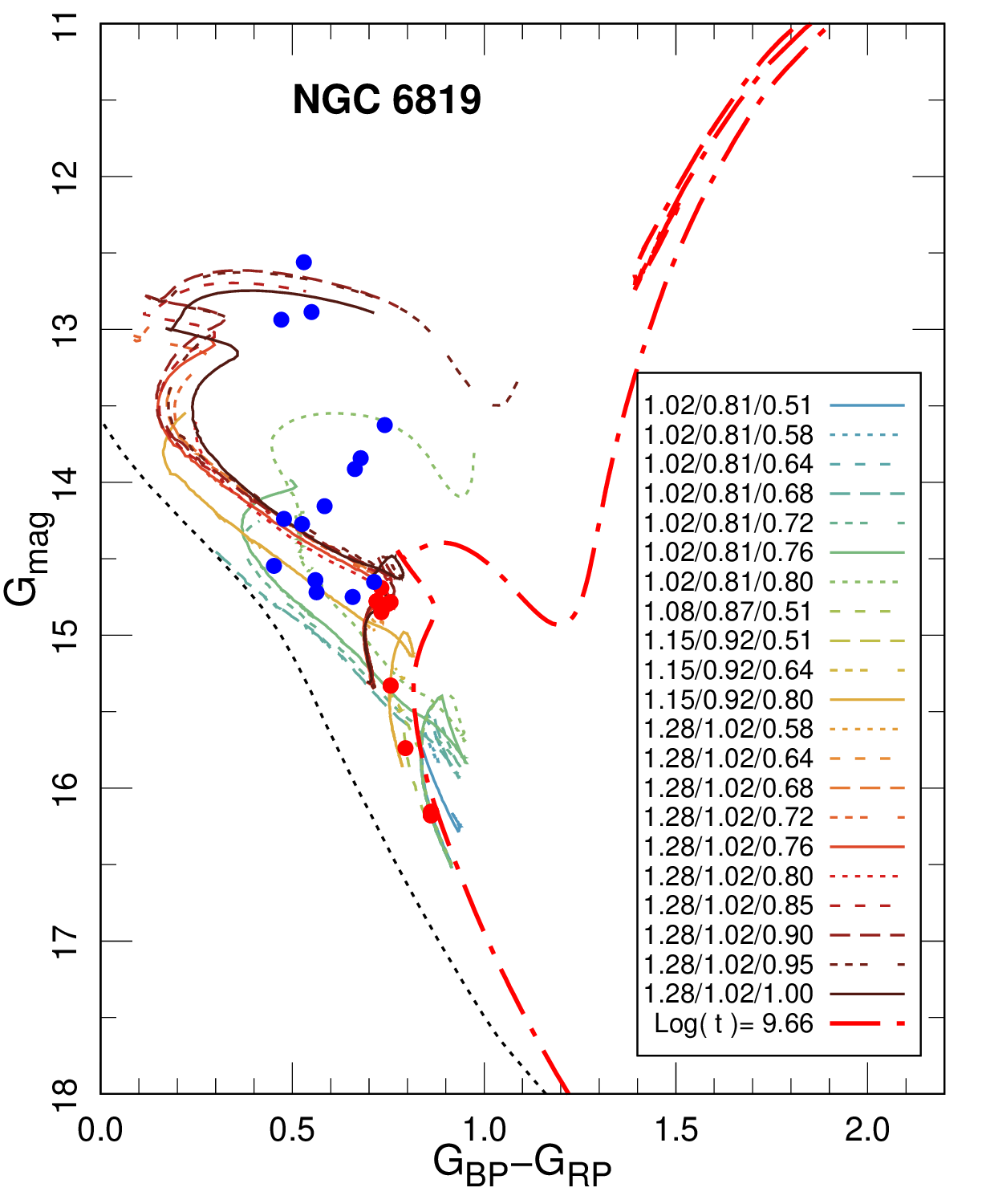}
\caption{Color-magnitude diagram for NGC~6819. We depict the combined binary tracks as solid, dotted, short-dashed, long-dashed, and three short-dashed colored lines. The dotted black line depicts the ZAMS, and the dot-dashed red line denotes the isochrone for NGC~6819 (in $Log(t[\mathrm{yr}])$) for single models. The filled blue circles indicate BSSs, and the filled red circles indicate the age on tracks. The three numbers in the box denote the $M_1$/$M_2$/$P$ donor mass (in $\mathrm{M}_{\sun}$), the accretor mass (in $\mathrm{M}_{\sun}$), and the period (in days).} \label{fig:NGC6819-bin}
\end{figure}

We computed the atmosphere models separately for each pair of binaries, that is, we obtained two sequences in the color-magnitude diagram for each binary. The emergent magnitude results from the sum of both fluxes, resulting in the composite evolutionary sequence. First, we computed the filter-weighted integrated fluxes $F_i$ for each model, 
\begin{equation}\label{eq_3}
F_{X,i}=\frac{\int f(\lambda) R_X(\lambda) d\lambda}{\int R_X(\lambda) d\lambda},
\end{equation}
where $i=A, B$. The letter A represents the donor star, and the letter B denotes the accretor star, $f(\lambda)$ is the computed absolute flux, and $R_X(\lambda)$ is the response function of filter $X$. In Equation~(\ref{eq_3}), the integration is over the filter bandpass $X$ in wavelength (angstroms), and the integrated flux $F_{X,i}$ is used to calculate magnitudes in each band by means of
\begin{equation}\label{eq_4}
M_{X,i}= -2.5 \log(F_{X,i}) + C_{X},
\end{equation}
where $M_{X,i}$ and $C_{X}$ are the synthetic magnitude and the flux constant for the X band, respectively. The flux constant is defined such that the synthetic magnitude matches the Vega magnitude in X band. However, when we observe the combined flux, then we should consider the following: 
\begin{equation}\nonumber
\frac{\int [ f_A(\lambda) + f_B(\lambda) ] R_X(\lambda) d\lambda}{\int R_X(\lambda) d\lambda}=F_{X,A} + F_{X,B}.
\end{equation}
Using Equation~(\ref{eq_4}), we obtain
\begin{equation}\label{eq_5}
10^{-0.4(M_{X,A}-C_X)}+10^{-0.4(M_{X,B}-C_X)}=F_{X,A} + F_{X,B}.
\end{equation}
When the binary system is not eclipsing,
\begin{equation}\label{eq_6}
M_{X,A+B}= -2.5 \log (F_{X,A} + F_{X,B}) + C_X,
\end{equation}
hence, 
\begin{displaymath}
M_{X,A+B}=
\end{displaymath}\begin{equation}\label{eq_7}
-2.5 \log(10^{0.4 C_X} [10^{-0.4 M_{X,A}}+10^{-0.4 M_{X,B}}]) + C_X.
\end{equation}
The sequences we show below were computed using the transformation given in Equation~(\ref{eq_7}) for the binary models listed in Table \ref{tab:binary_models}.\\

In Fig.~(\ref{fig:NGC6819-bin}), we depict the transformed binary sequences. %\LEt{figure details such as line styles should not be repeated in the main text. To my understanding, the next sentence ought to be removed as well. Please check this and apply it throughout}Blue filled-circles represent the expected BSSs members of NGC~6819.

\section{Cluster-by-cluster discussion}\label{sec:comparison}
In this section, we comment on the comparison between binary star models and the distribution of stragglers in the Gaia color-magnitude diagrams.
Before entering the cluster-by-cluster analysis, we stress that the comparison is qualitative at this stage because we lack crucial information about basic straggler properties such as mass and period. The obvious aim of this comparison is to highlight that binary evolution models can fit the general distribution of BSSs in the cluster CMDs well.

\subsection{NGC 6819}
The comparison is illustrated in Figure \ref{fig:NGC6819-bin}. The distribution of bona fide 
BSS members (filled blue circles) is compared with evolutionary tracks for the mass ratios indicated in the inset. For reference, the cluster best-fit isochrone is drawn with dashed-dotted line. %\LEt{do you mean "lines"?}. 
This cluster harbors BSSs  with very different luminosity, and they seem to form two groups. A bright group of three BSSs comes from the binary evolution of pairs with high-mass donors (1.28 $M_{\odot}$), while the fainter group follows the evolutionary path of pairs with lower-mass donors (1.02 $M_{\odot}$). Overall, the binary tracks match the BSS distribution excellently. Some BSSs are in the stage of leaving their MS according to the models.

\subsection{Berkeley~32}
In this old open cluster, most BSSs are crowded immediately above the TO. However, a number of BSSs is much brighter than the TO, and they cover a large range in magnitude (up to 5 magnitudes; see Figures \ref{fig:Berkeley_32} and \ref{fig:Berkeley32_b}). The cluster hosts two YSSs as well. The large excursion in magnitude (and hence in mass) implies that very different binary evolution tracks have to be employed to cover the BSS region. We note that both BSSs and YSSs are well fit by these models, however.
Finally, a few faint stragglers are not reproduced by the suite of models we ran. These faint stars may be misclassified stragglers. Another possibility is that different parameters of the binary model need to be adopted.

\subsection{Berkeley~39}
The location of BSS candidates in the CMD of this cluster is shown in Figures \ref{fig:Berkeley_39} and \ref{fig:Berkeley39_b}. As noted for other cases (see below), the BSS candidates are separated into two main groups. One group consists of faint BSSs that closely follow the low-mass donor tracks. The other group contains BSSs that are spread in color and is significantly redder than the ZAMS. These are well matched by the binary star evolutionary models with a donor mass in the range $1.28-1.60 M_{\odot}$ overimposed in the CMD. Berkeley~39 also contains two YSSs.
As in the case of Berkeley~32 and given our stringent and conservative selection criteria, some of these stars can be misclassifications and/or it could be necessary to consider different parameters for the binary model. 

\subsection{Collinder~261}
This cluster harbors a rich population of BSS candidates. They compose a continuous luminosity sequence that overlaps either with the ZAMS or with binary evolution sequences. We note that several BSSs lie below and redward of the cluster TO. If these two BSSs are confirmed, they might be two examples of the so-called sub-subdwarfs \citep{Geller_2017b, Geller_2017a}. Another interesting case is the bright BSS immediately above the cluster TO, which looks like a promising YSS candidate, but fails to follow the standard criterion for the location of these stars in the CMD. The very faint BSS candidate at G $\sim$ 17.2 might be a misidentification, or might more conservatively be off the ZAMS because of larger photometric errors. CMDs of Collinder~261 can be found in Figures \ref{fig:Collinder261} and  \ref{fig:Collinder261_b}.

\subsection{Melotte~66}
The BSSs for this cluster seem to form a double sequence (see Figures \ref{fig:Melotte66} and \ref{fig:Melotte66_b}). A first sequence follows a low-mass donor ($\sim 1.02 M_{\odot}$) tracks and is relatively close to the cluster ZAMS. 
A second redder sequence is also tightly aligned with the binary tracks. It has a pair with a 1.28 $M_{\odot}$ donor. Double sequences have widely been discussed in GCs, and many authors claimed that they can be used as tracers of a core-collapse event and subsequent dynamical evolutionary phases in GCs.  NGC~6256 \citep{Cadelano_2022}, NGC362 \citep{Dalessandro_2013}, and NGC~1261 \citep{Simunovic_2014} are notable examples for which authors have explored these double sequences.  The red BSS sequence is interpreted as populated by MT products \citep{Xin_2015}, and the blue narrow sequence may be reproduced by collisions or merger products \citep{Sills_2009}.
{This is not universal, however, as W UMa contact binaries were discovered in the blue and red sequence of the GC M30 \cite{Ferraro_2009b}. A few studies (e.g., \citealt{Stepien_2015, Jiang_2017}) indicated that a binary evolution contributes to the formation of blue stragglers in both sequences.
As in our case, \cite{Rao_2023} found that this feature is present in OCs. These authors claimed, more conservatively, that MT causes the formation of both sequences. 
Furthermore, \cite{Rao_2022} performed a multiwavelength spectral energy distribution (SED) fitting for 12 out of the 18 BSSs of Melotte~66, and only 2 of them (BSS3, Gaia DR3 5507234395259864448, and BSS6, 5507234429619602944) were identified as MT products. All the remaining stragglers are consistent with a single-component SED. Interestingly enough, the two binary stars fall within the blue sequence. 
Unfortunately, given the lack of studies, no other variable stars have been reported within the Melotte~66 blue straggler population. Additional spectroscopic observations or light-curve modeling is required to confirm the binary nature of the remaining 16 BSSs and to constrain its spectroscopic parameters, rotational velocity, mass, and so on. An analysis of the individual nature of blue stragglers and the origin of the apparent double sequences in Melotte~66 is far beyond the scope of this paper, however.
Additionally, we confirm that single YSS candidate (yellow dot) is interpreted as an evolved binary system. 

\subsection{NGC~188}
Like M~67 (see below), this cluster has been studied very extensively, and its population of BSSs is quite well established. Their distribution is shown in Figures \ref{fig:NGC188} and \ref{fig:NGC188_b}. The BSS region is crossed by several binary star evolutionary tracks, which confirm the interpretation that BBS candidates far from the ZAMS are the result of some type of binary star evolution.

\subsection{NGC~2141}
This rich cluster hosts a significant number of BSS candidates (see Figures \ref{fig:NGC2141} and \ref{fig:NGC2141_b}). We cannot exclude that some of them are a misidentification, in particular, those very close to the cluster sequence and TO. In any case, the bulk of these BBS is well matched by the binary evolution tracks for the masses indicated in the inset. Moreover, the YSS (yellow dot) is  quite well reproduced.

\subsection{NGC~2158}
The distribution of BSS candidates is shown in Figures \ref{fig:NGC2158} and \ref{fig:NGC2158_b}. A significant number of BSS candidates lies blueward of the ZAMS.  This might be a misidentification or the product of photometric errors. We conservatively considered them to be single BSSs that follow the cluster ZAMS. The distribution of all the other BSS candidates is well reproduced by the set of binary star evolutionary models, including the one yellow straggler.

\subsection{NGC~2243}
This old open cluster (Figures \ref{fig:NGC2243} and \ref{fig:NGC2243_b}) has only a few BSSs. As seen in other cases, the BSS candidates that do not lie along the ZAMS (five to six, in this case) are well matched by binary evolution tracks. Some of them (two to three) also appear to be evolved binary systems, but they are not yet in the so-called YSS region according to the standard definition.

\subsection{NGC~2506}
This cluster is illustrated in Figure \ref{fig:NGC2506}. With the exception of just one BSS that lies at the top of the ZAMS, all the candidate BSSs follow the pattern of binary star evolution covering a sizeable luminosity range. The BSSs are not very numerous, as in the case of NGC~2682 (see below), but this clearly depends on the cluster mass. No YSSs are detected.

\subsection{NGC~2682}
The population of BSSs in M67 was frequently studied in the past. Several of these stars have indeed been identified as binaries. This is confirmed by the overall agreement between BSS candidates and the binary tracks in Figures\ref{fig:NGC2682_b}. The cluster also harbors two YSSs that seem to be evolved binary systems (see Figure \ref{fig:NGC2682_b}).

\subsection{NGC~7789}
NGC~7789, like Melotte~66, is a rich and compact old open cluster. 
Although less evident than in Melotte~66, a double sequence of BSSs seems to be present in NGC~7789 as well (see Figures \ref{fig:NGC7789} and \ref{fig:NGC7789_b}). The two sequences are almost parallel, one follows low-mass donor ($1.28 M_{\odot}$) tracks, the other follows the MS of the binary tracks with parameters 1.60, 1.28, and 1.25.  
Additionally, employing a time-series radial velocity analysis, the study by \cite{Nine_2020} revealed binary characteristics of four BSSs. Moreover, in a separate work, \cite{Vaidya_2022b} presented the SED of five BSS candidates, confirming the presence of a hot companion for all five. Two of these stars overlap with the sample identified by \cite{Nine_2020}. The remaining eight BSSs did not exhibit any type of variability. Upon comparing their findings with our own, we observed that all BSSs, regardless of whether they are binary or single, exhibit a random distribution in the CMD.

\section{Summary}\label{sec:conclusions}
The aim of this work was twofold.\\
On the one hand, we focused on estimating the binary fraction in a sample of BSSs hosting old open clusters to assess the relation between the population of BSSs and the parent cluster binary fraction.
On the other hand, we compared the distribution of BSSs in the cluster CMD  with the evolutionary track of binary star evolution at varying mass ratio.

Using the Python unsupervised algorithm \texttt{pyUPMASK} and the automated stellar cluster analysis (\texttt{ASteCA}) package, we identified cluster members, cluster parameters, and blue straggler populations. We estimated the binary fraction, mass, distance, and age for a total of 12 open clusters, and we investigated their possible relation with the observed number of BSSs, $\mathrm{N_{BSS}}$. Finally, we compared the distribution of the straggler population in the color-magnitude diagram with a set of evolutionary tracks derived from binary evolution models.

Our research provided direct evidence that clusters within our regime produce BSSs more efficiently.  Furthermore, we confirmed that the number of BSSs is influenced by the binary fraction ($\mathrm{f_{bin}}$), mass, and the number of binaries (N$_{\mathrm{bin}}$), and that the correlation with the binary fraction is the strongest predictor of the blue straggler population size with a Spearman coefficient of $\mathrm{r_{s}}\sim0.84$, followed by the number of binaries and the total mass, with $\mathrm{r_{s}}=0.80$ and $\mathrm{r_{s}}=0.58$, respectively.
Additionally, as in the case of the GCs, we found that the correlation between the stragglers and N$_{\mathrm{bin}}$ improves the previously known sublinear correlation between the straggler numbers and the mass of the cluster.

We also estimated the dependence of the straggler size and the different cluster parameters. We found a dependence of the form (in decreasing order) $\mathrm{N_{BSS}} \propto \mathrm{N_{bin}^{0.75\pm0.13}}$, $\mathrm{N_{BSS}} \propto \mathrm{M_{Tot}^{0.6\pm0.2}}$, and $\mathrm{N_{BSS}} \propto \mathrm{f_{bin}^{0.5\pm0.11}}$ for the number of binaries, the total cluster mass, and the binary fraction, respectively. 

On the other hand, having established these observational pieces of evidence, we employed evolutionary models of binary star evolution to explore the multidimensional parameter of the mass ratio, primary mass, and period.  Lacking  precise estimates of the BSS period, the exploration at this stage was rather qualitative.
We still showed that for all the clusters under investigation, the binary evolution tracks can reproduce the bulk of the BSSs distribution in the CMDs well.
One of the most interesting outcomes is the evidence of a dichotomy in the BSSs distribution in a few clusters. We identified a sequence of stragglers lying close to the cluster ZAMS, while the other sequence follows the binary evolution tracks better. This is similar to the BSSs distribution that was recently found in some globular clusters.
We cannot argue, however, that this is a general feature because we do not see it clearly in all clusters, possibly because of photometric errors or an uncertain membership.
We are confident that with more precise observational data, in particular, with the knowledge of the binary period, a better comparison can be performed in the future and that the properties of several stragglers can be revealed.
An additional improvement is certainly to search for a more precise binary mass ratio for each of these clusters. Our assumed mass ratio of 0.7 does not produce a perfect fit for all the clusters we studied.

With pleasure, we thank the anonymous referee for the many valuable suggestions and comments, which improved the paper greatly.

\begin{acknowledgements}
\\
The work of M.J.~Rain and G. Carraro has been supported by Padova University grant BIRD191235/19: {\it Internal dynamics of Galactic star clusters in the Gaia era: binaries, blue stragglers, and their effect in estimating dynamical masses}.\\
O. G. Benvenuto is member of the Carrera del Investigador Cient\'{\i}fico, Comisi\'on de Investigaciones Cient\'{\i}ficas de la Provincia de Buenos Aires, Argentina.\\
S. Villanova gratefully acknowledges the support provided by Fondecyt regular n. 1220264 and by the ANID BASAL projects FB210003.
\end{acknowledgements}

\bibliography{bib}
\bibliographystyle{aa}
%\begin{thebibliography}{}
%\end{thebibliography}

%
%-------------------------------------------------------------
%          For the appendices, table longer than a single page
%-------------------------------------------------------------

% Table will be print automatically at the end of the document, 
% after the whole appendices

\begin{appendix} %First appendix
\onecolumn 
\section{Observational color-magnitude diagrams}\label{append_sec:cmds}
%\vspace*{+1.2cm}
%\counterwithin{figure}{section}
\begin{figure*}[htp]
%\vspace*{-0.25cm}
\begin{subfigure}[b]{0.5\textwidth}
\centering
\includegraphics[width=0.8\textwidth]{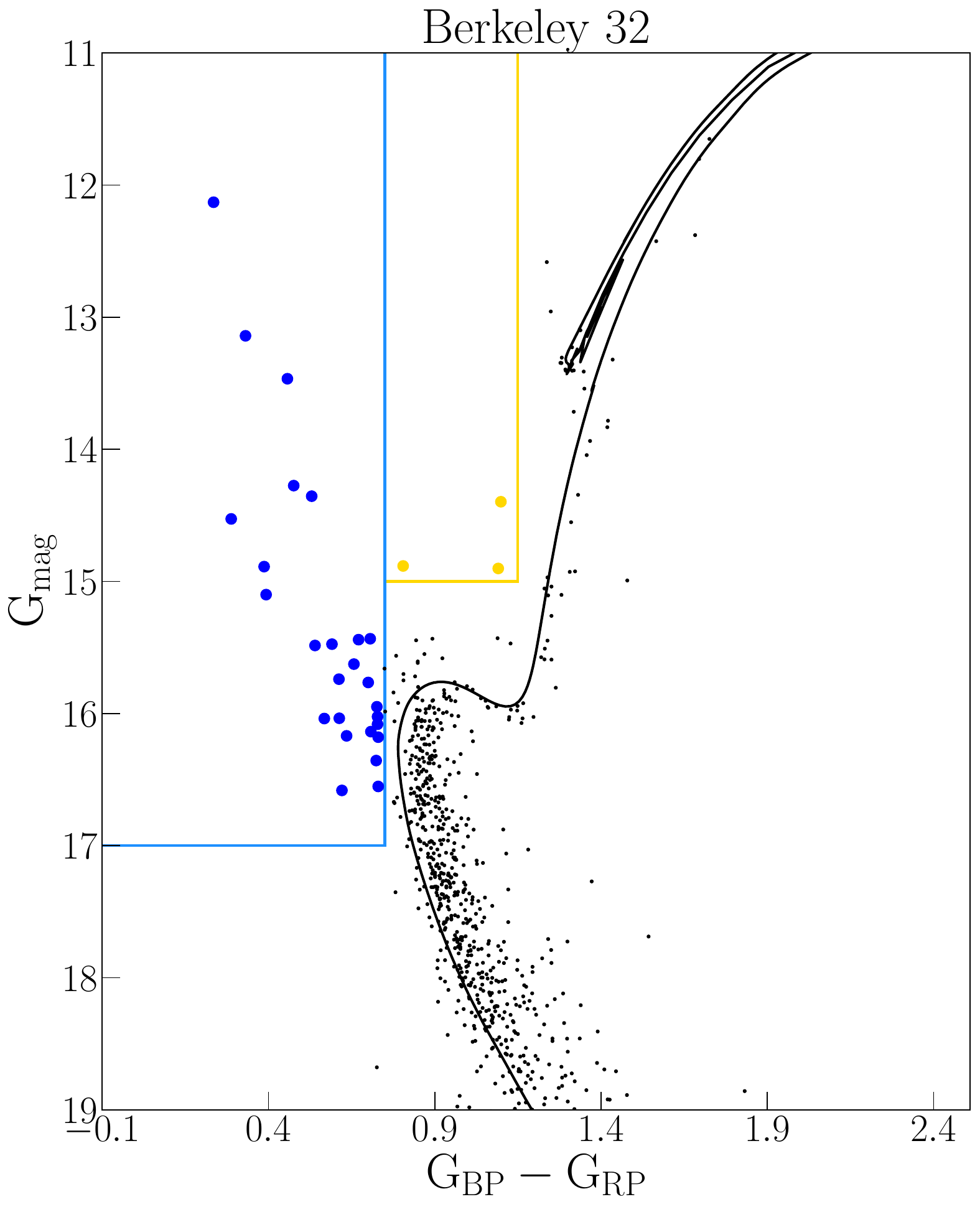}
\caption{}
\label{fig:Berkeley_32}
\end{subfigure}
\hfill
\begin{subfigure}[b]{0.5\textwidth}
\centering
\includegraphics[width=0.8\textwidth]{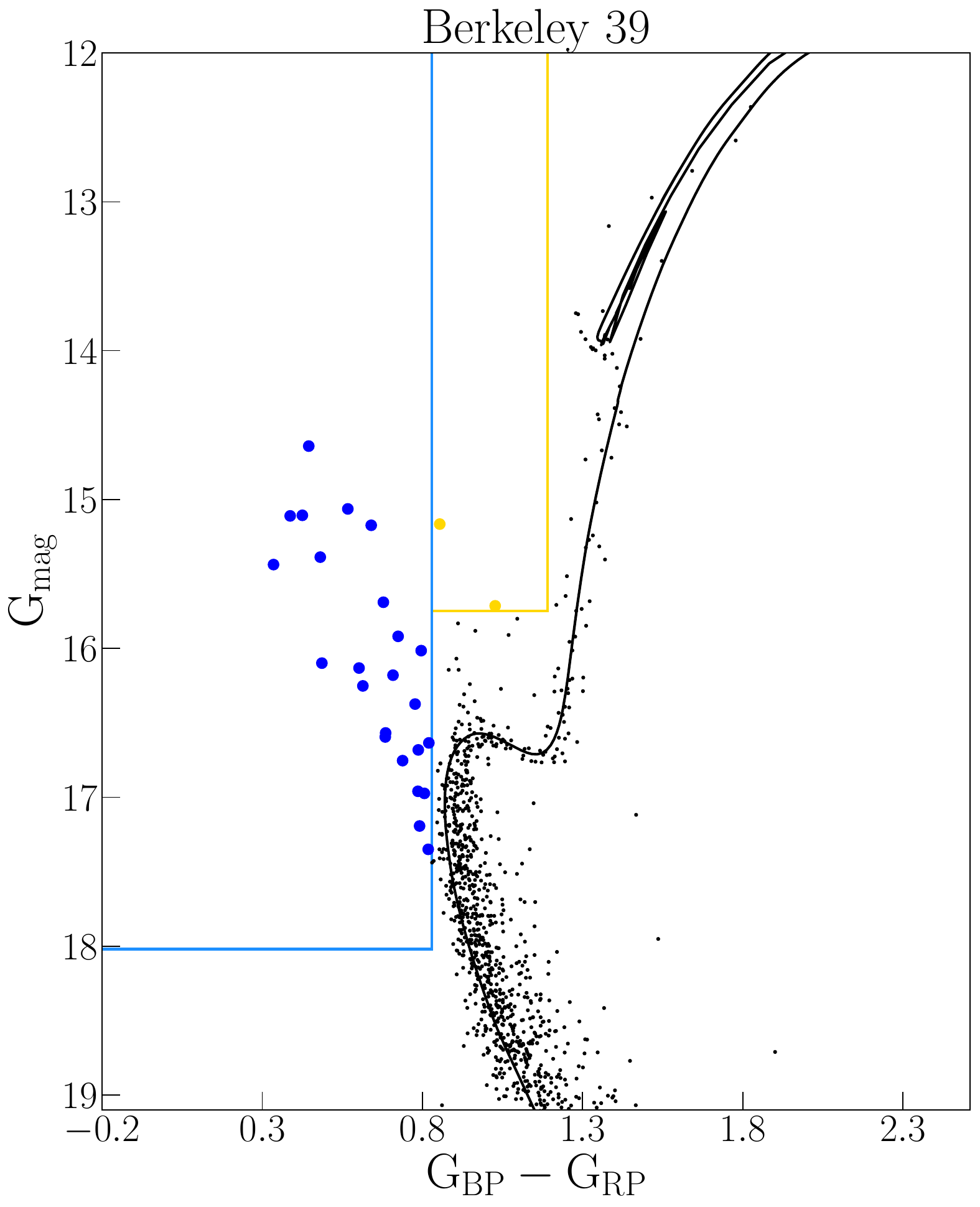}
\caption{}
\label{fig:Berkeley_39}
\end{subfigure}
\hfill
\begin{subfigure}[b]{0.5\textwidth}
\centering
\includegraphics[width=0.8\textwidth]{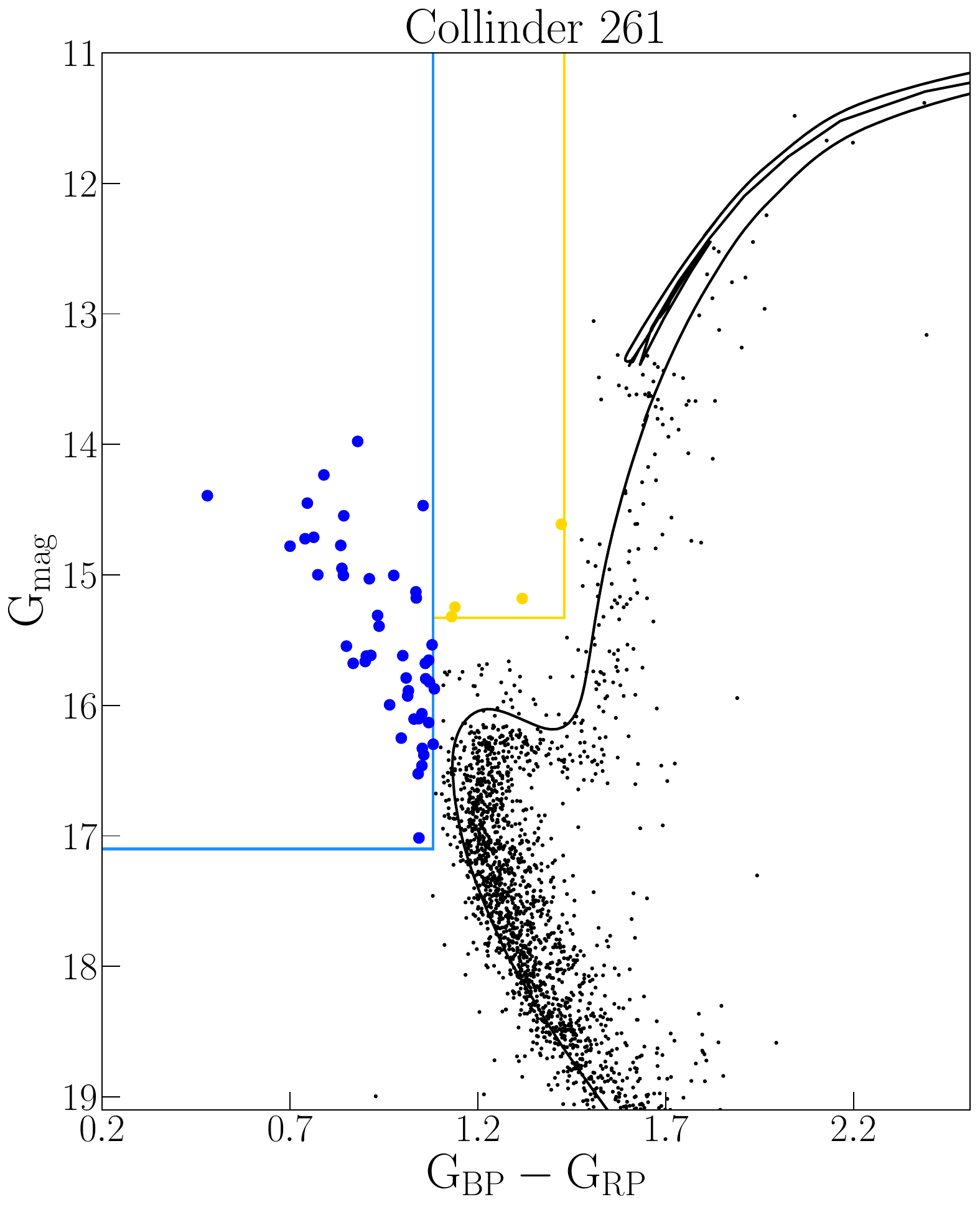}
\caption{}
\label{fig:Collinder261}
\end{subfigure}
\hfill
\begin{subfigure}[b]{0.5\textwidth}
\centering
\includegraphics[width=0.8\textwidth]{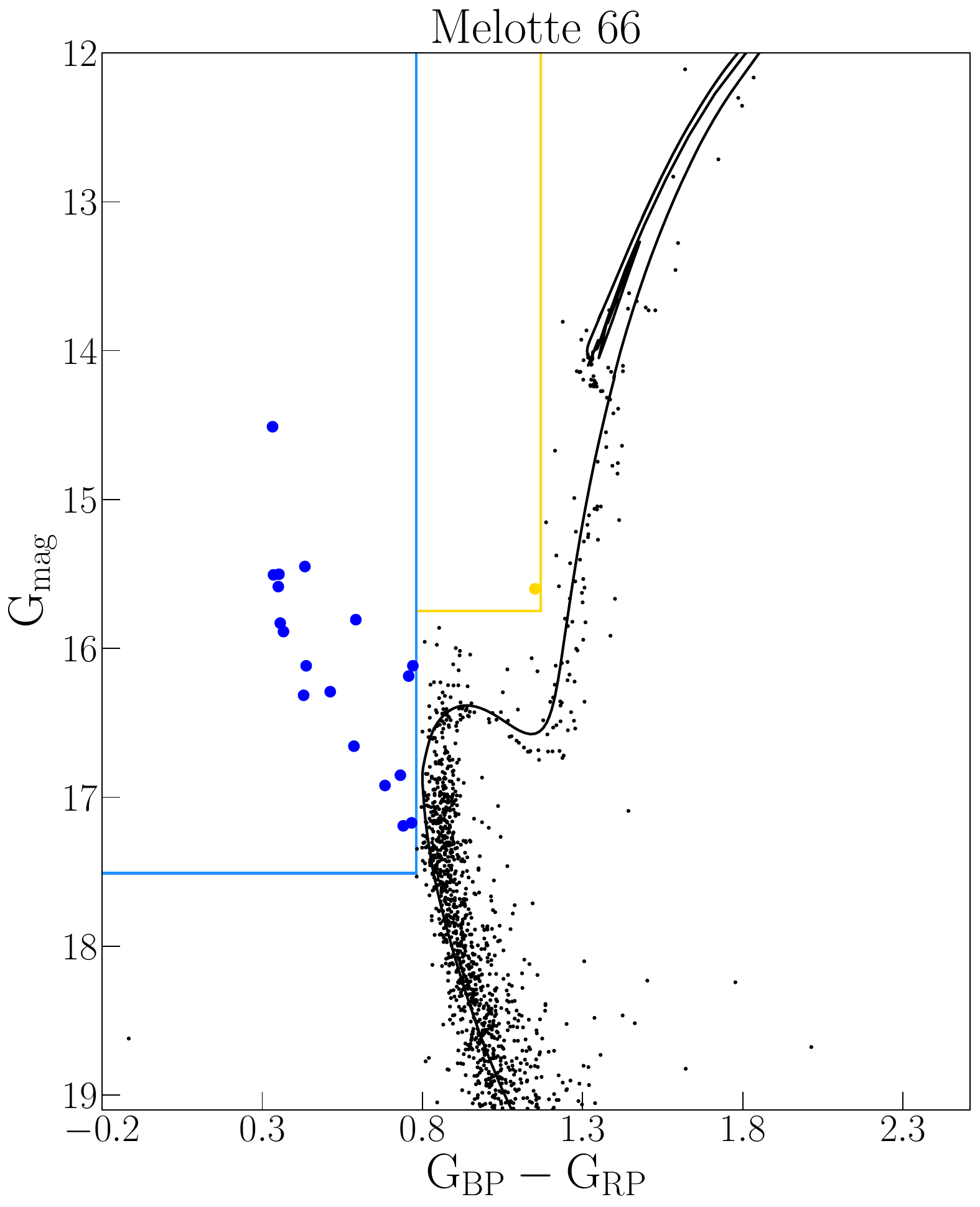}
\caption{}
\label{fig:Melotte66}
\end{subfigure}
\hfill
\caption{Color-magnitude diagrams of all 11 remaining clusters in our sample.  The filled black circles are the {\it~Gaia} members selected as described in \S~\ref{sec:data_selection}. The filled blue and yellow circles are the blue and yellow straggler stars selected as described in \S~\ref{section:bs_ys_region}. The black line shows the corresponding isochrone  \citep{Dotter_2016} with the age, Av, and [Fe/H] value of each cluster.} \label{fig:cmd_others}
\end{figure*}

\begin{figure*}[htp]
\begin{subfigure}[b]{0.5\textwidth}
\centering
\includegraphics[width=0.8\textwidth]{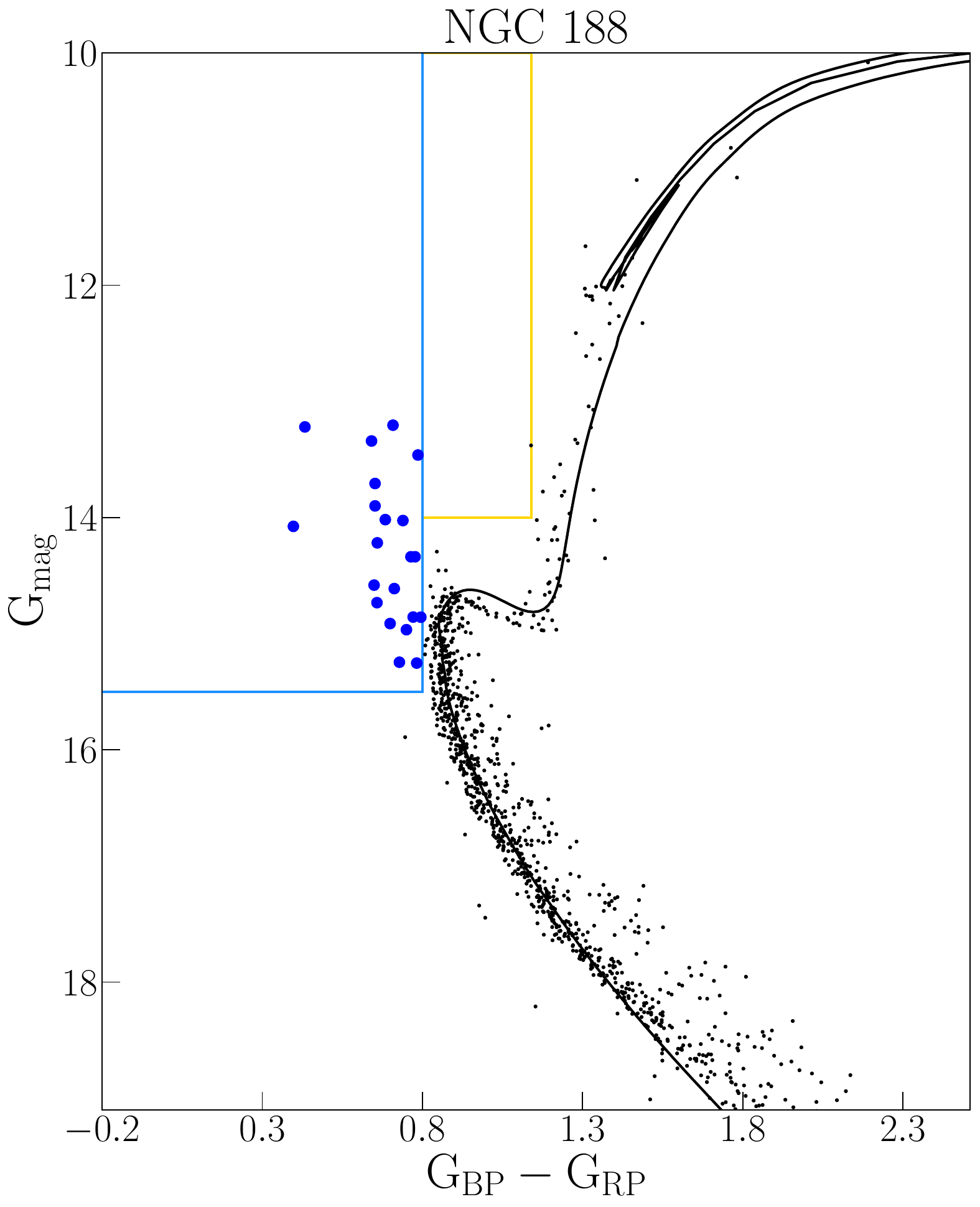}
\caption{}
\label{fig:NGC188}
\end{subfigure}
\hfill
\begin{subfigure}[b]{0.5\textwidth}
\centering
\includegraphics[width=0.8\textwidth]{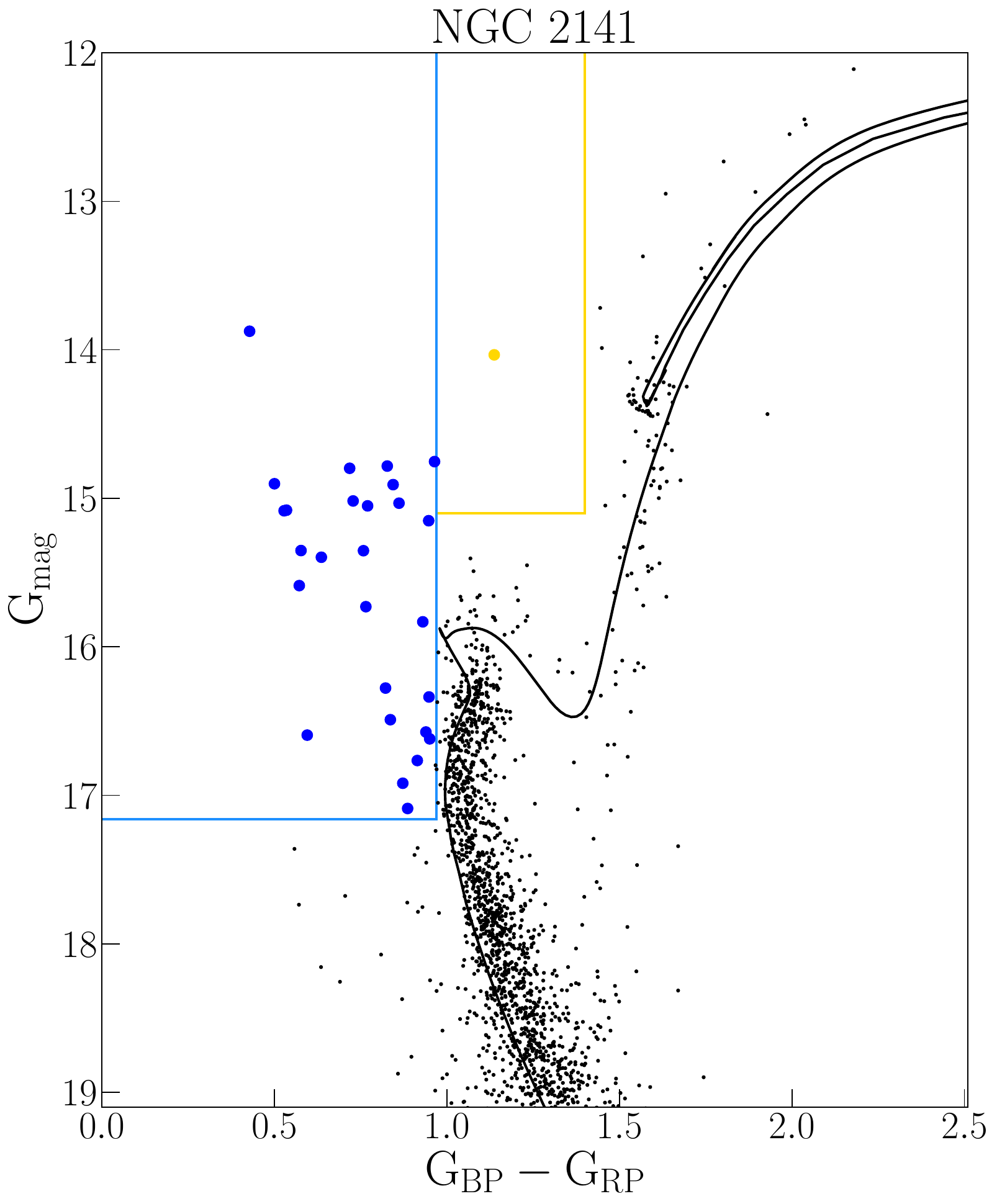}
\caption{}
\label{fig:NGC2141}
\end{subfigure}
\hfill
\begin{subfigure}[b]{0.5\textwidth}
\centering
\includegraphics[width=0.8\textwidth]{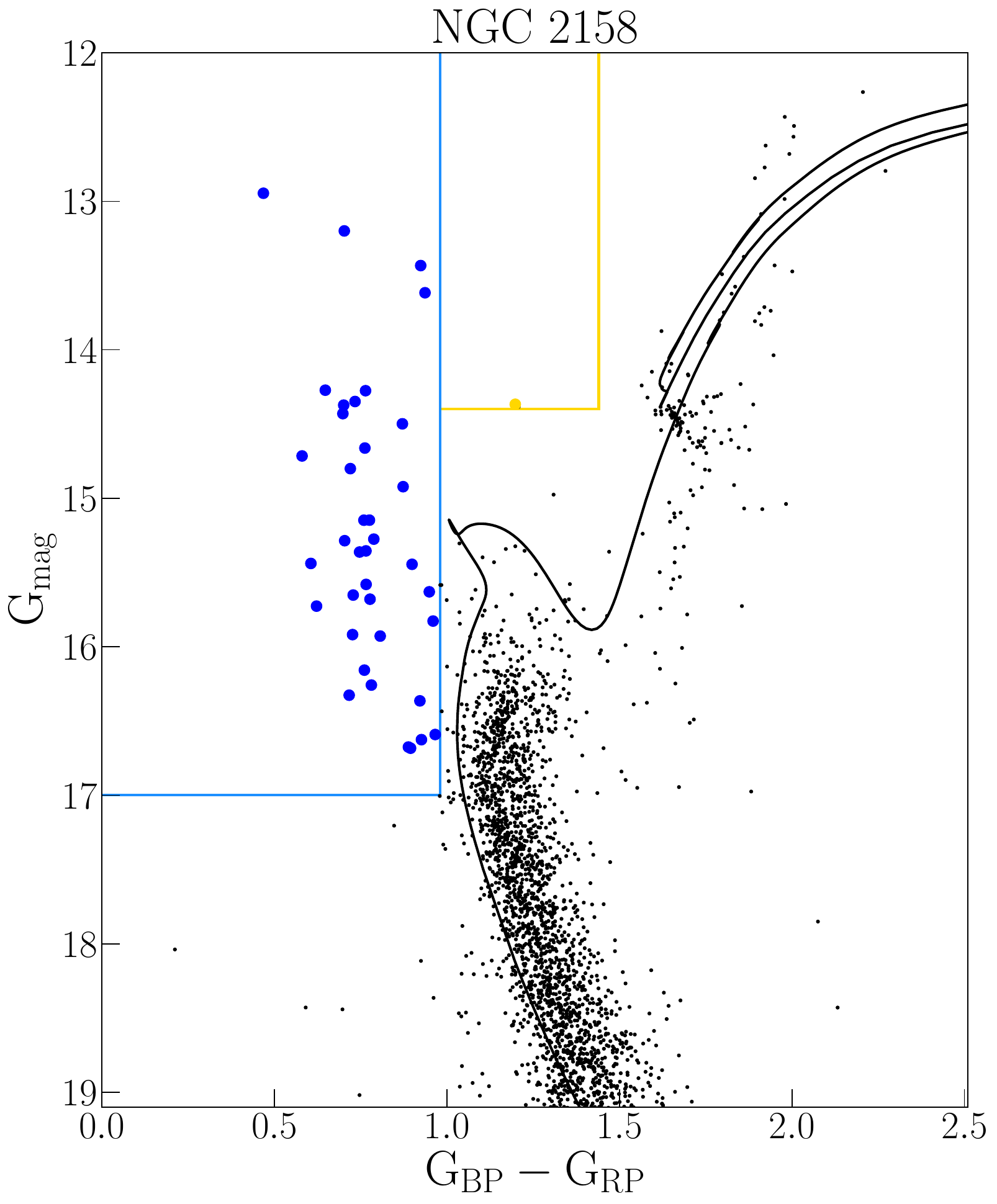}
\caption{}
\label{fig:NGC2158}
\end{subfigure}
\hfill
\begin{subfigure}[b]{0.5\textwidth}
\centering
\includegraphics[width=0.8\textwidth]{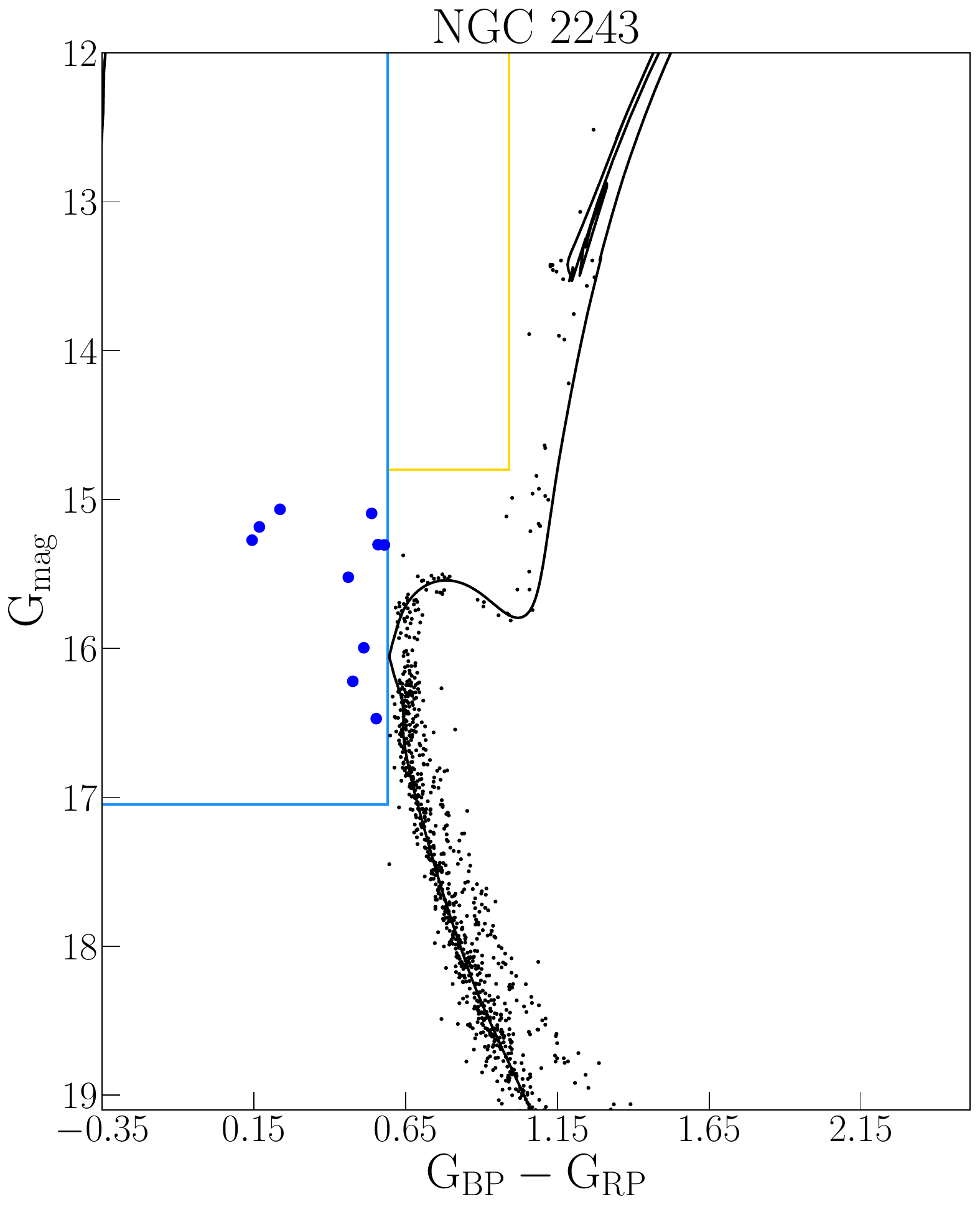}
\caption{}
\label{fig:NGC2243}
\end{subfigure}
\caption{Same as. Fig.~\ref{fig:cmd_others}.}
\end{figure*}

\vspace*{+6cm}\begin{figure*}[htp]
\centering
\begin{subfigure}[b]{0.49\textwidth}
\centering
\includegraphics[width=0.81\textwidth]{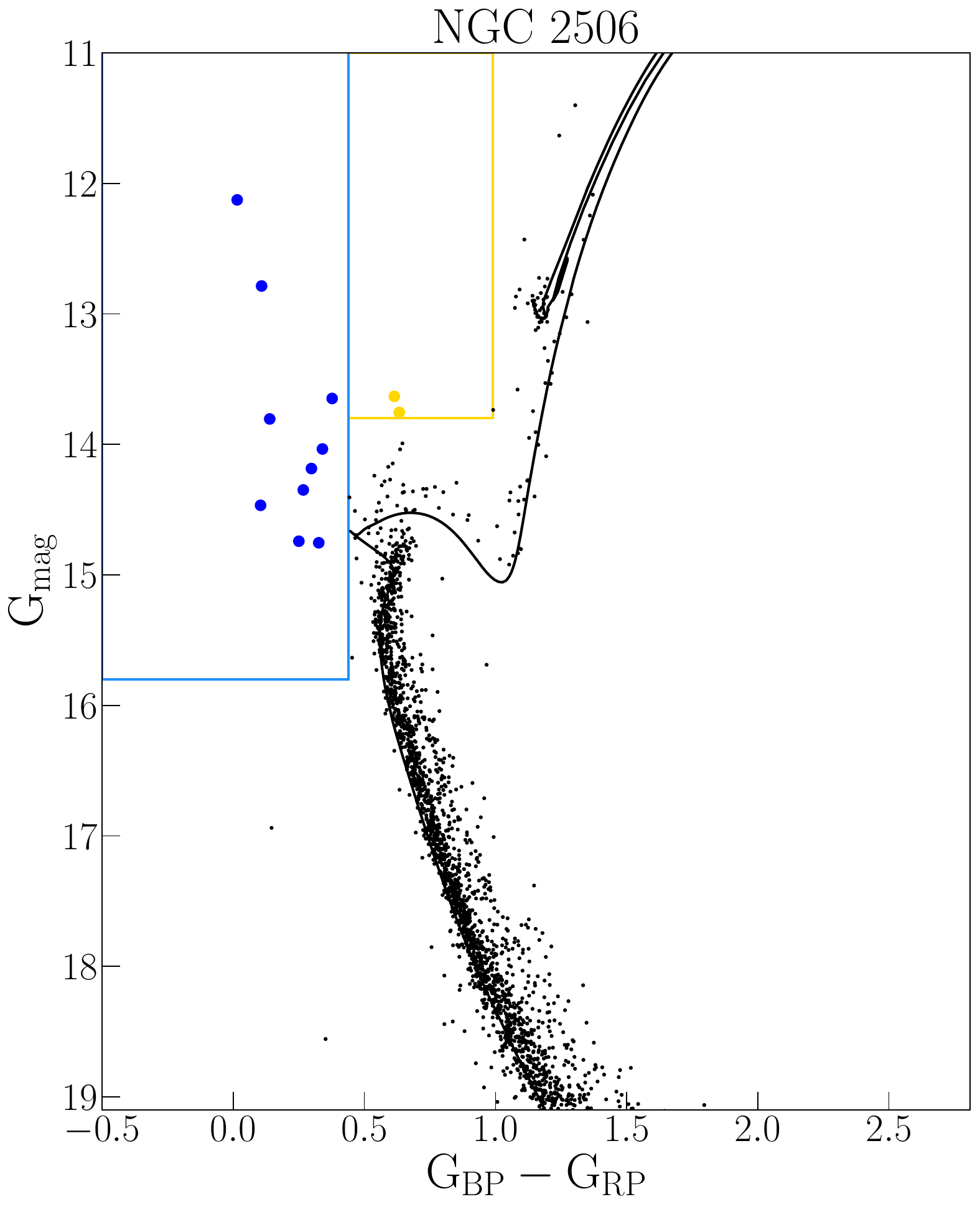}
\caption{}
\label{fig:NGC2506}
\end{subfigure}
%\hfill
\begin{subfigure}[b]{0.5\textwidth}
\centering
\includegraphics[width=0.8\textwidth]{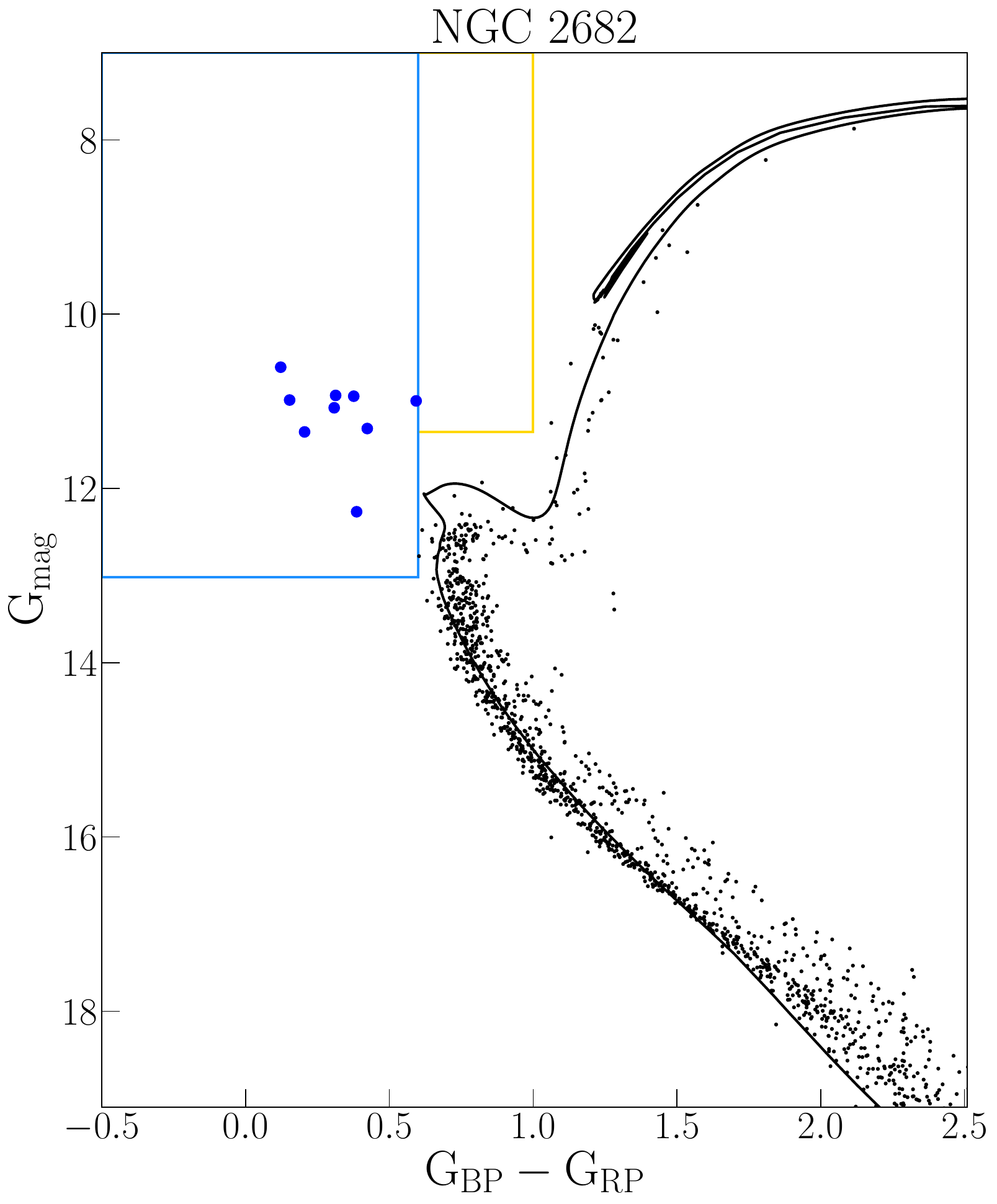}
\caption{}
\label{fig:NGC2682}
\end{subfigure}
%\hfill
\begin{subfigure}[b]{0.5\textwidth}
\centering
\includegraphics[width=0.8\textwidth]{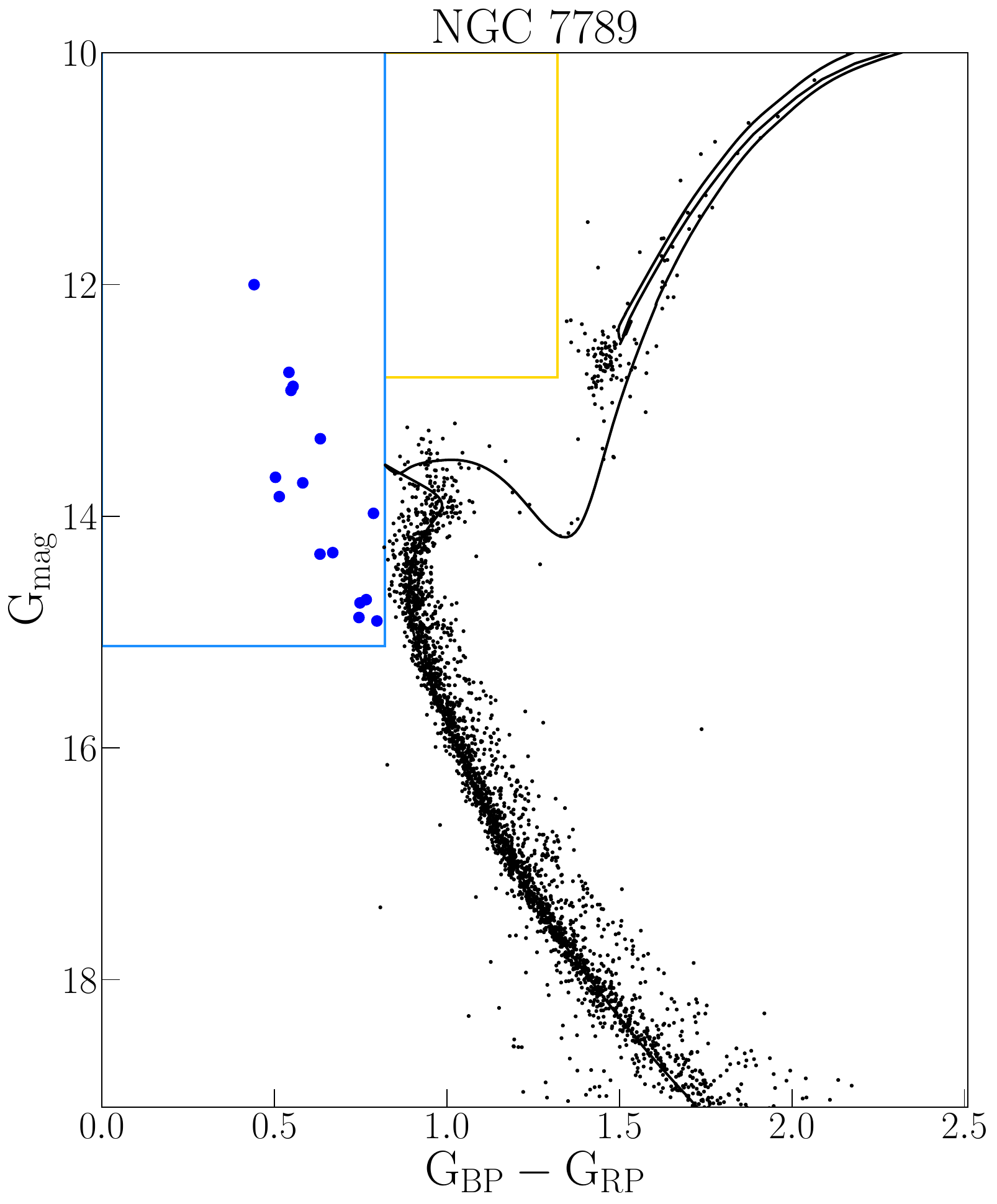}
\caption{}
\label{fig:NGC7789}
\end{subfigure}
\caption{Same as. Fig.~\ref{fig:cmd_others}.}
\end{figure*}

\vspace*{+20cm}
\section{Theoretical color-magnitude diagrams}\label{append_sec:hrdgaia}
\begin{figure*}[htp]
\begin{subfigure}[b]{0.5\textwidth}
\centering
\includegraphics[width=\textwidth,  angle =270 ]{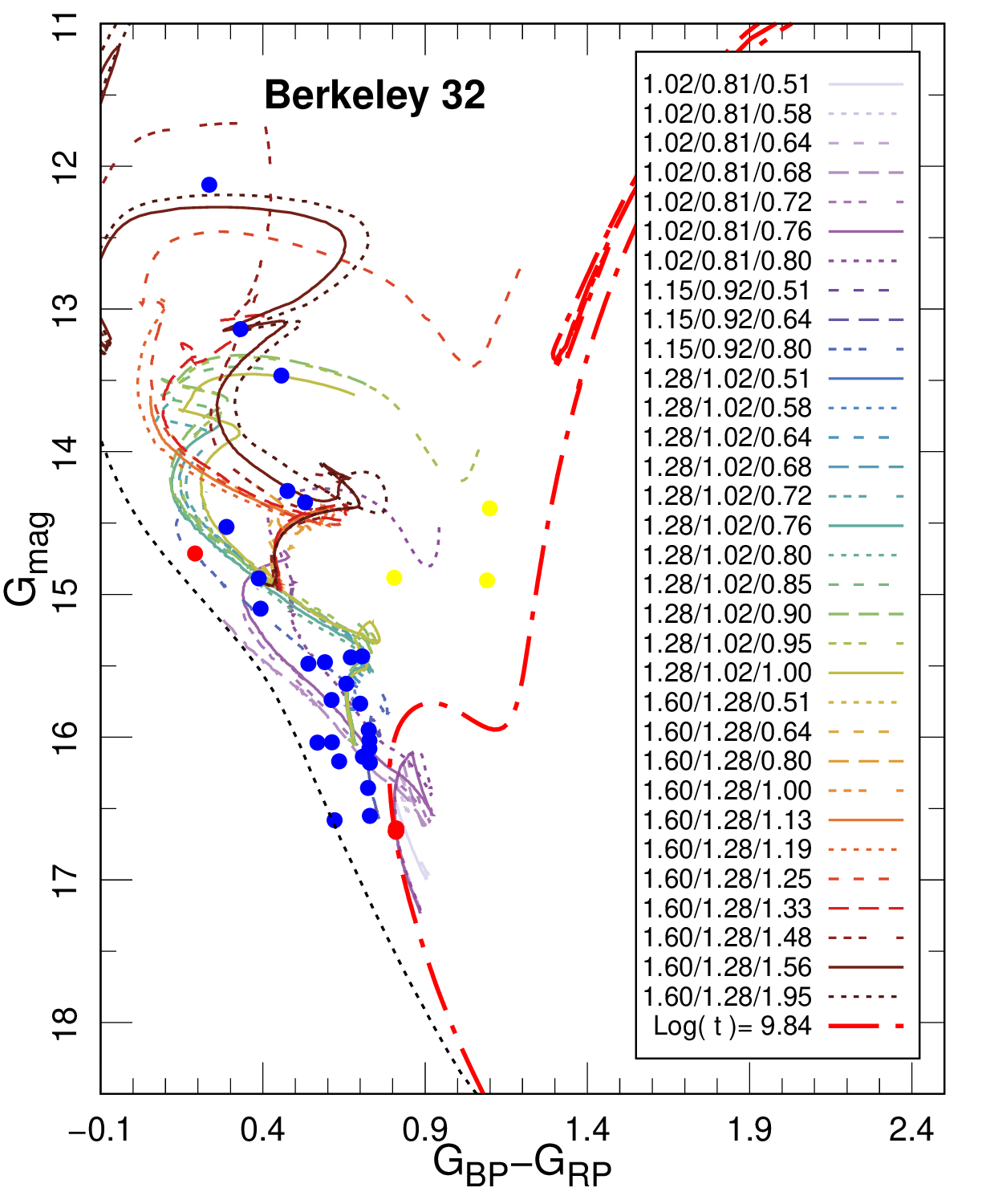}
\caption{}
\label{fig:Berkeley32_b}
\end{subfigure}
\begin{subfigure}[b]{0.5\textwidth}
\centering
\includegraphics[width=\textwidth,  angle =270 ]{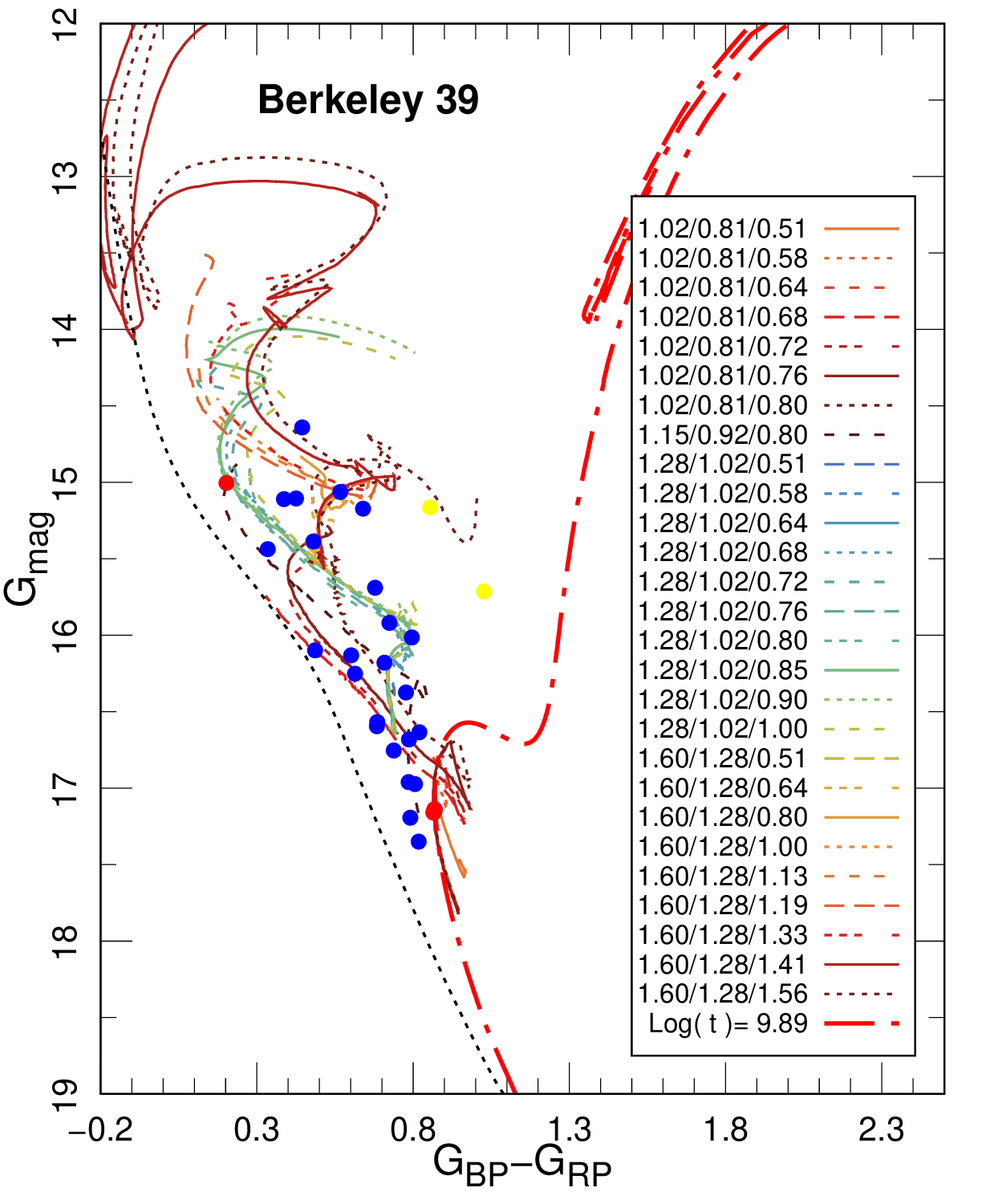}
\caption{}
\label{fig:Berkeley39_b}
\end{subfigure}
\begin{subfigure}[b]{0.5\textwidth}
\centering
\includegraphics[width=\textwidth,  angle =270 ]{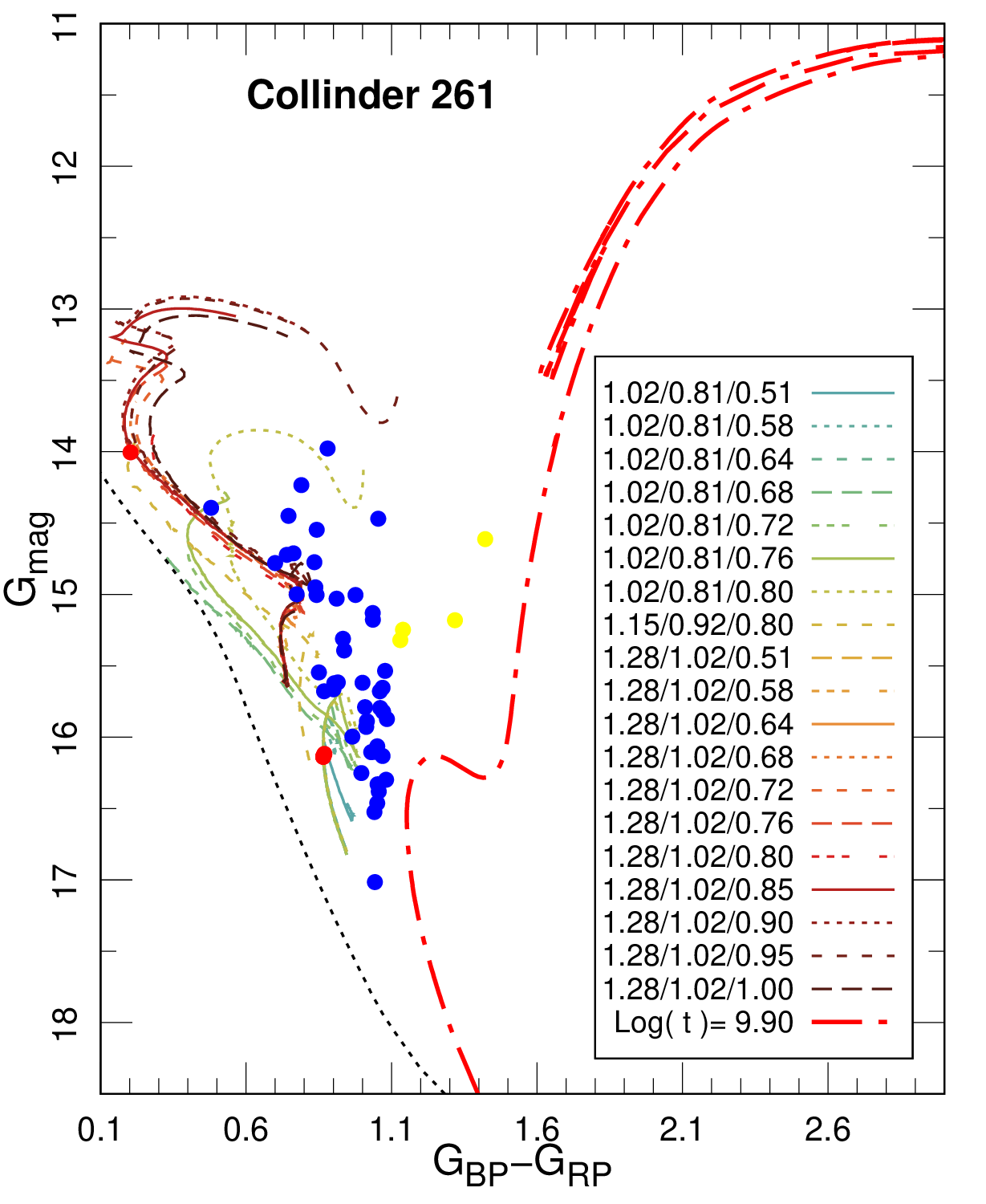}
\caption{}
\label{fig:Collinder261_b}
\end{subfigure}
\begin{subfigure}[b]{0.5\textwidth}
\centering
\includegraphics[width=\textwidth,  angle =270 ]{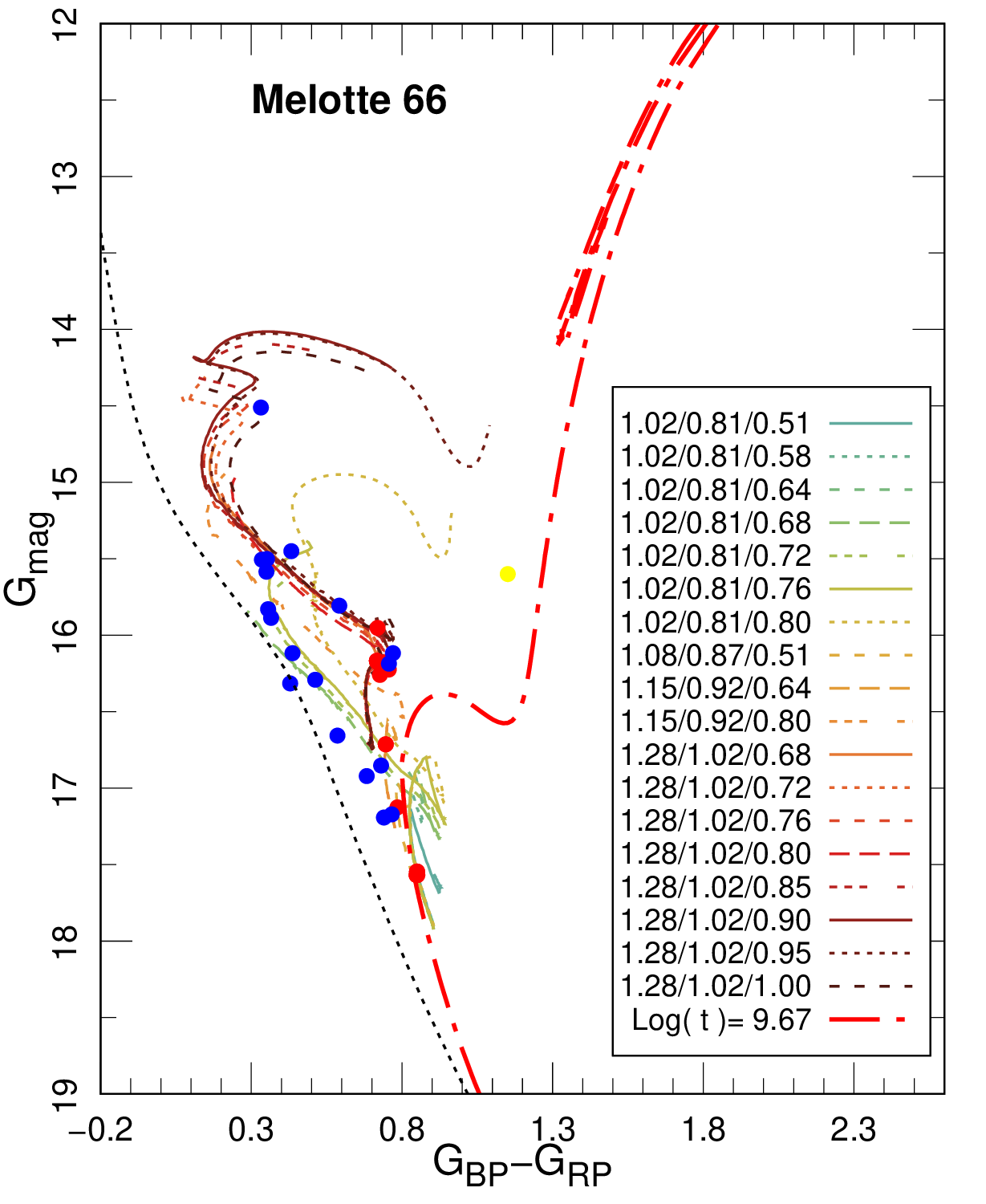}
\caption{}
\label{fig:Melotte66_b}
\end{subfigure}
\caption{Color-magnitude diagrams. The combined binary tracks are represented as solid, dotted, short-dashed, long-dashed, and three short-dashed colored lines. The dotted black lines depict the ZAMS, and dot-dashed red lines represent the corresponding isochrones (in $Log(t[\mathrm{yr}])$) for single models. The BSSs are indicated by filled blue circles, and YSSs are shown by filled yellow circles. {The filled red circles indicate the cluster age.} The sets of three numbers denote the  $M_1$/$M_2$/$P$ donor mass (in $\mathrm{M}_{\sun}$), the accretor mass (in $\mathrm{M}_{\sun}$), and the period (in days).} \label{fig:six_gaia_hrd1}
\end{figure*}

\vspace*{+1cm}\begin{figure*}[htp]
\begin{subfigure}[b]{0.5\textwidth}
\centering
\includegraphics[width=\textwidth,  angle =270 ]{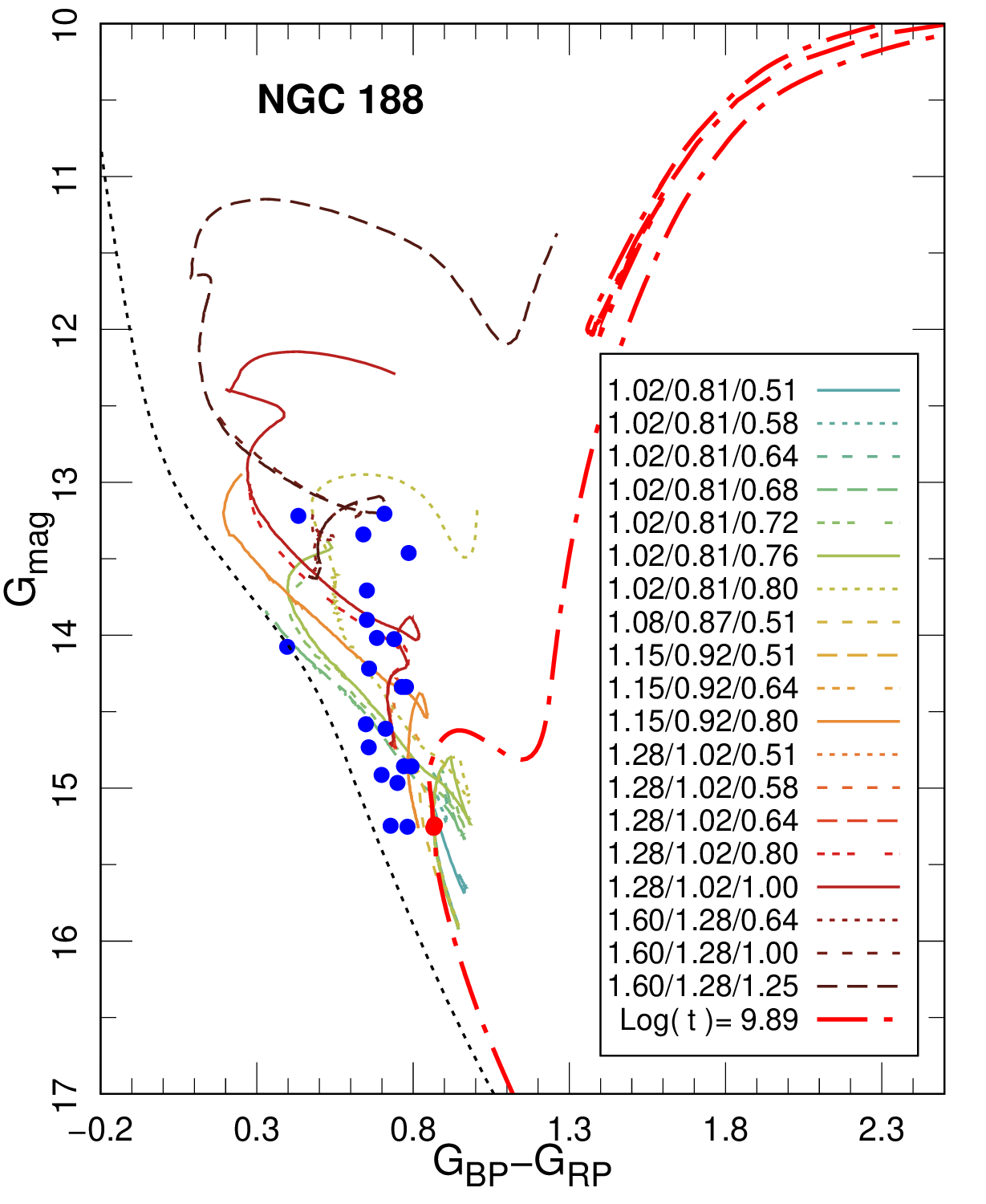}
\caption{}
\label{fig:NGC188_b}
\end{subfigure}
\begin{subfigure}[b]{0.5\textwidth}
\centering
\includegraphics[width=\textwidth,  angle =270 ]{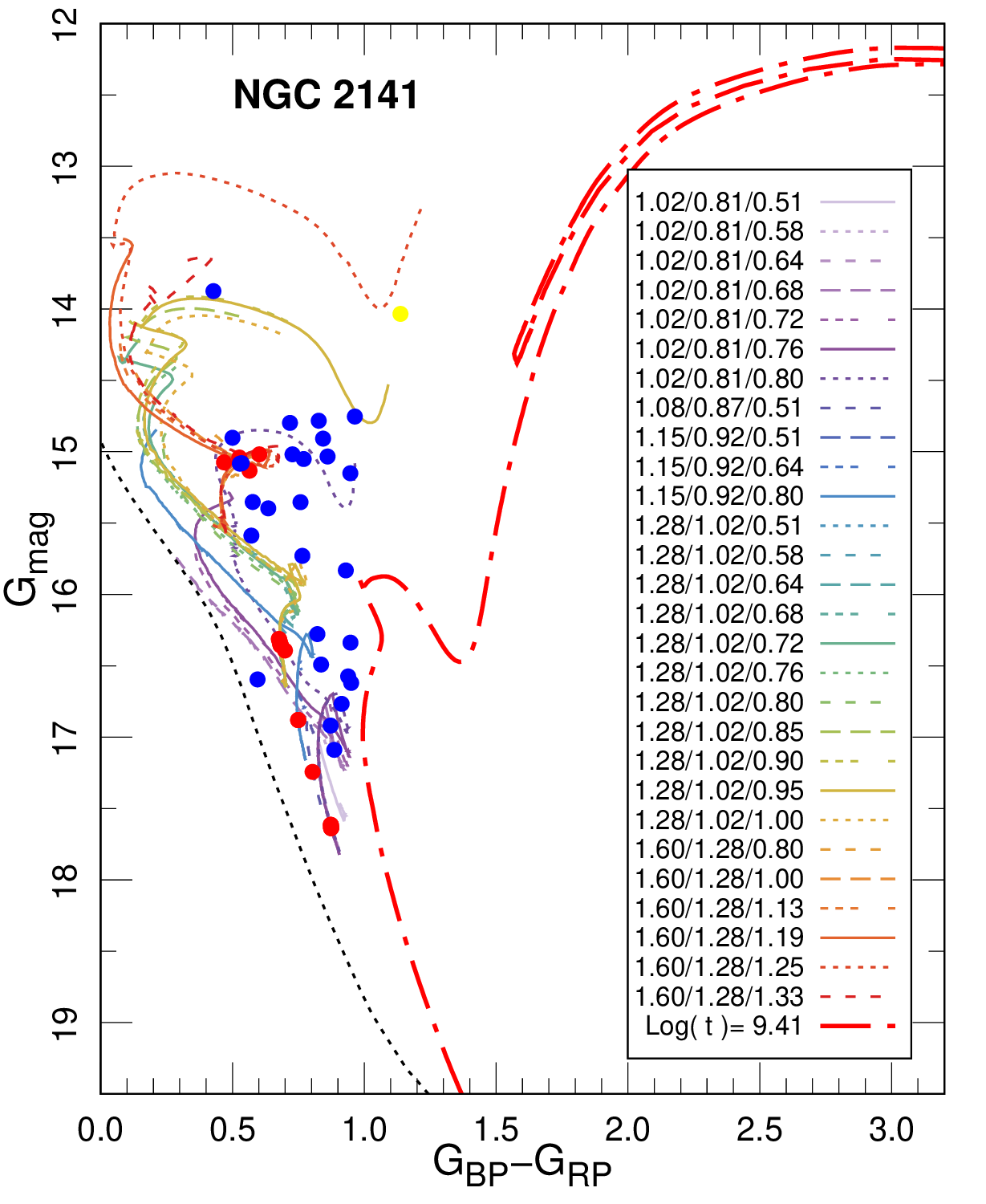}
\caption{}
\label{fig:NGC2141_b}
\end{subfigure}
\begin{subfigure}[b]{0.5\textwidth}
\centering
\includegraphics[width=\textwidth,  angle =270 ]{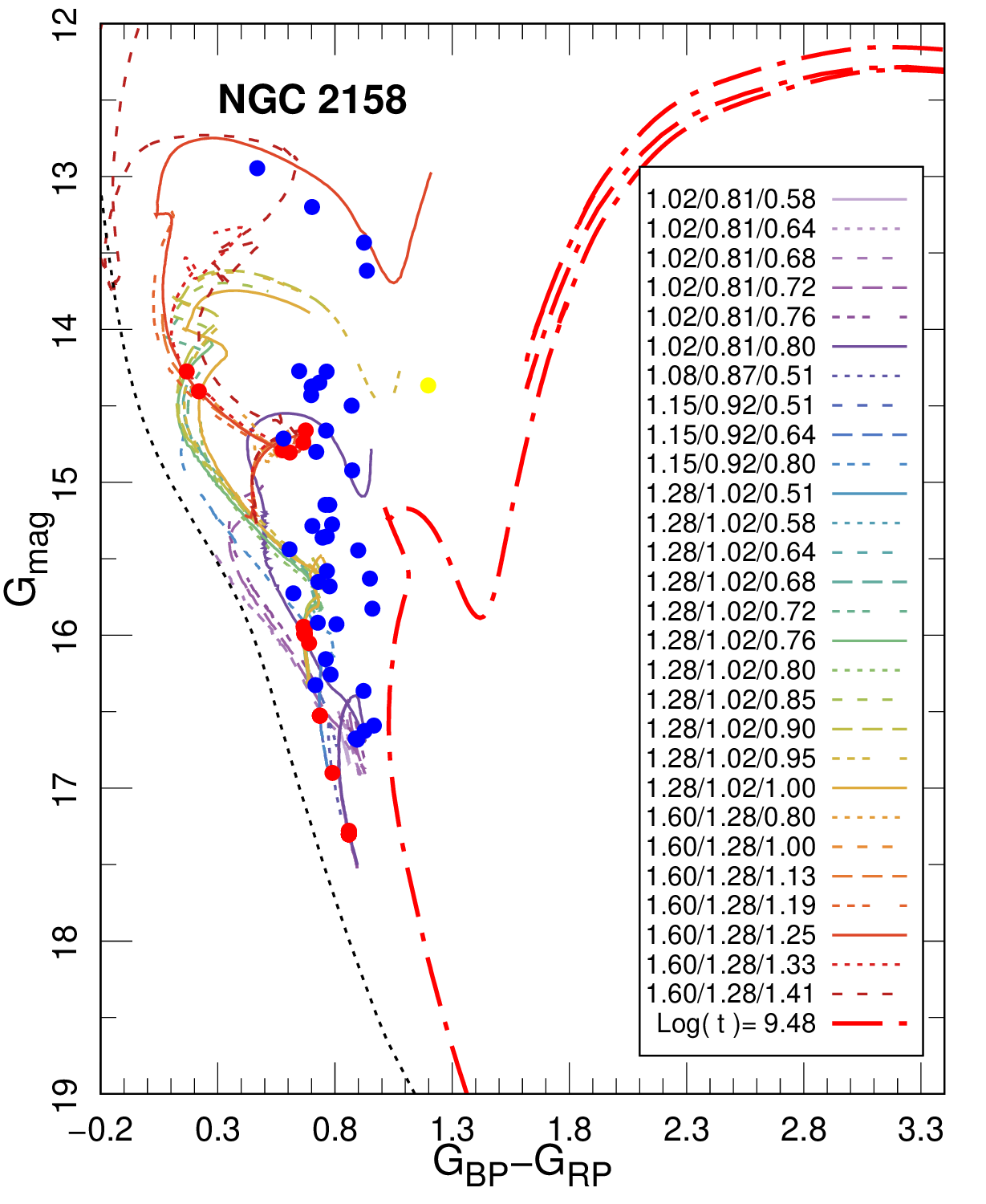}
\caption{}
\label{fig:NGC2158_b}
\end{subfigure}
\begin{subfigure}[b]{0.5\textwidth}
\centering
\includegraphics[width=\textwidth,  angle =270 ]{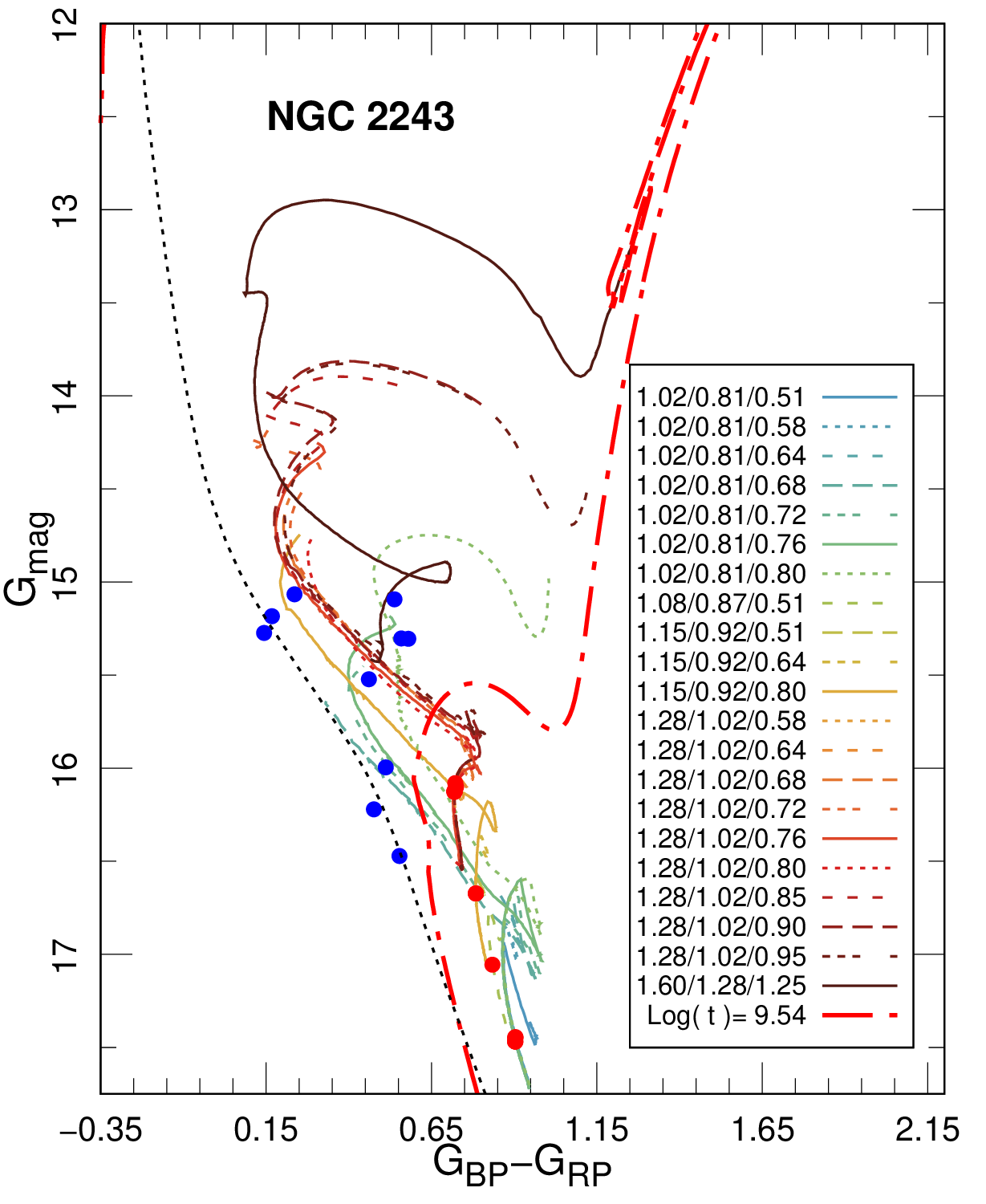}
\caption{}
\label{fig:NGC2243_b}
\end{subfigure}
\caption{Same as Fig.~\ref{fig:six_gaia_hrd1}.}
\end{figure*}

\vspace*{+2cm}\begin{figure*}[htp]
\begin{subfigure}[b]{0.5\textwidth}
\centering
\includegraphics[width=\textwidth,  angle =270 ]{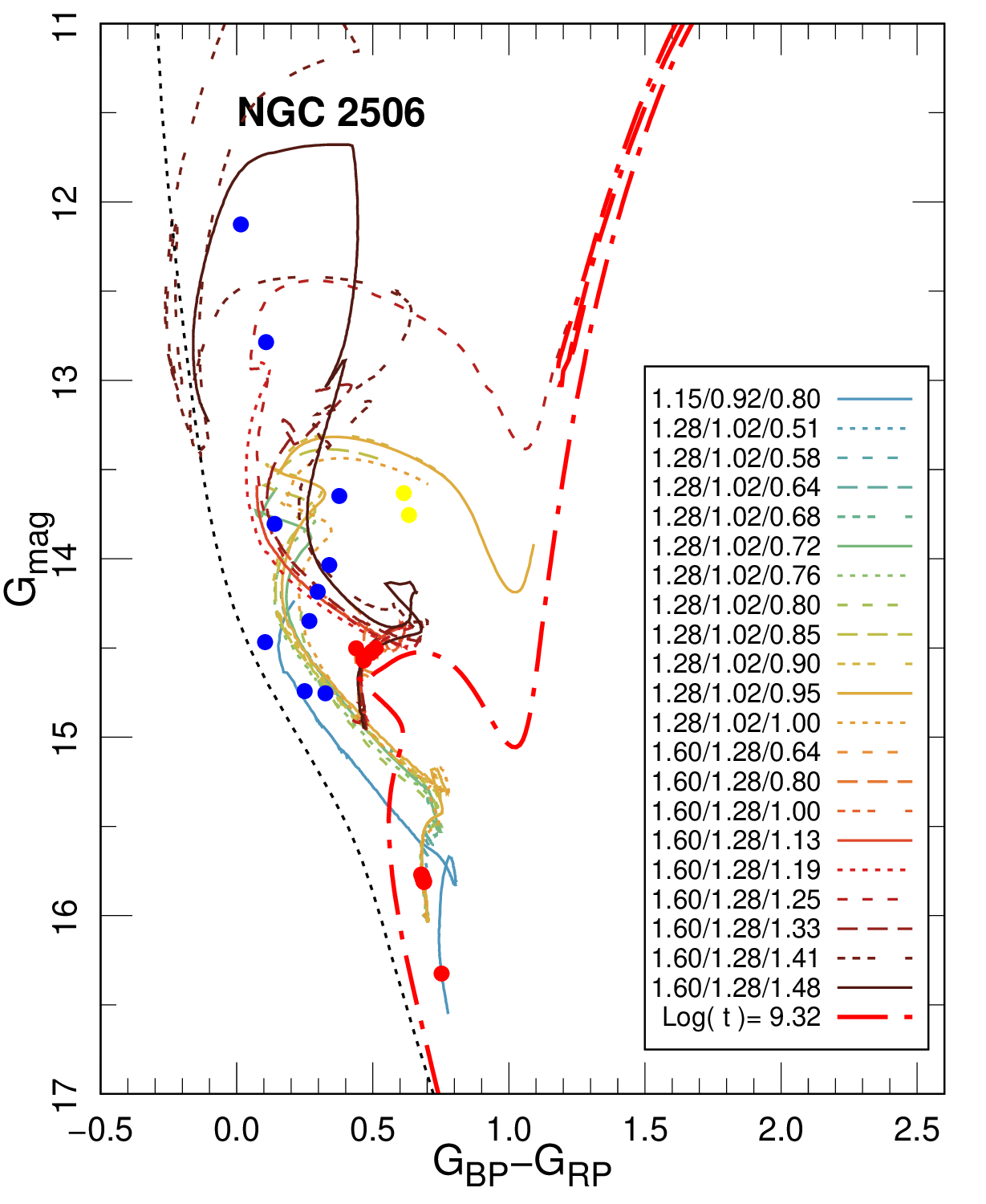}
\caption{}
\label{fig:NGC2506_b}
\end{subfigure}
\begin{subfigure}[b]{0.5\textwidth}
\centering
\includegraphics[width=\textwidth,  angle =270 ]{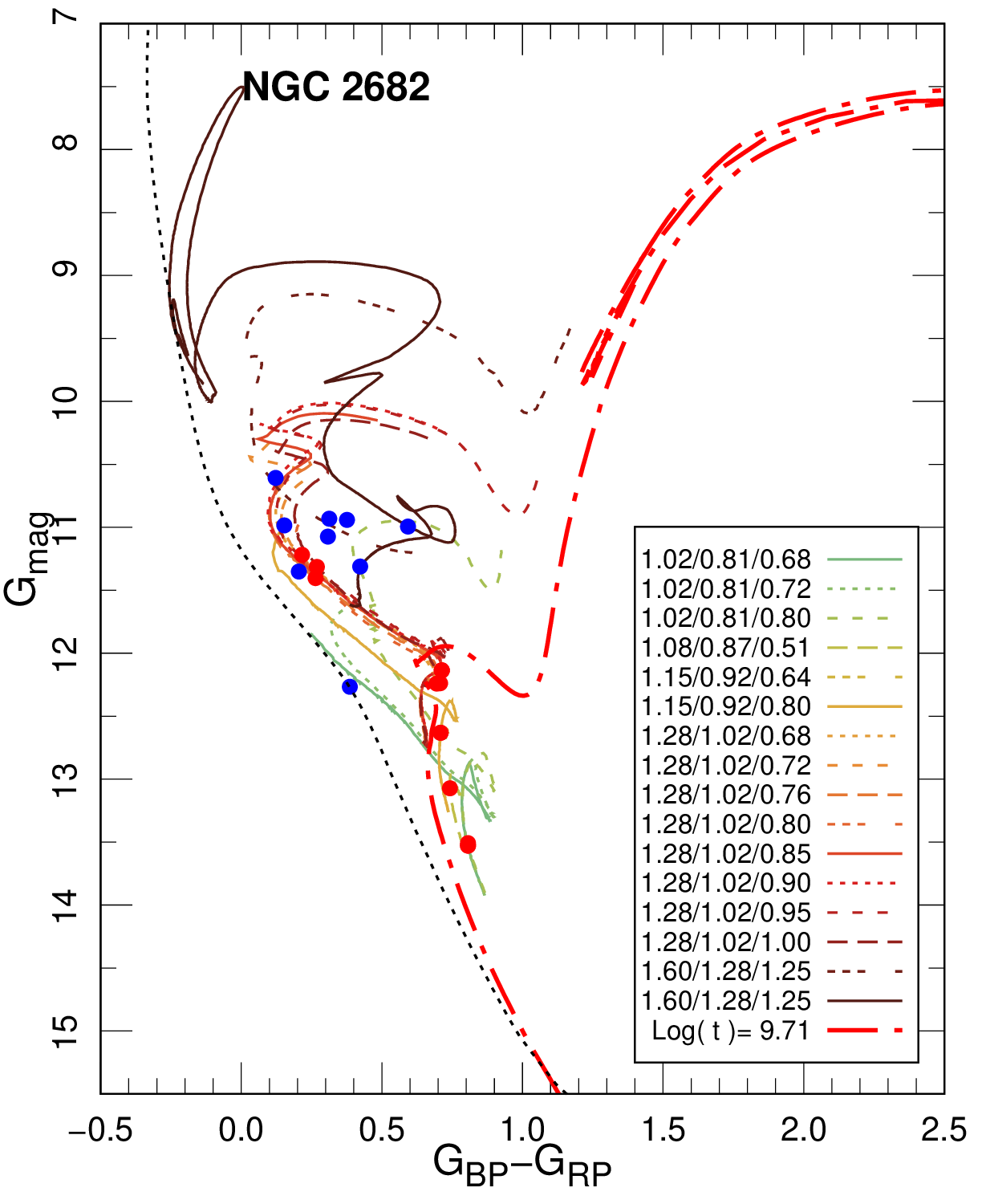}
\caption{}
\label{fig:NGC2682_b}
\end{subfigure}
\begin{subfigure}[b]{0.5\textwidth}
\centering
\includegraphics[width=\textwidth,  angle =270 ]{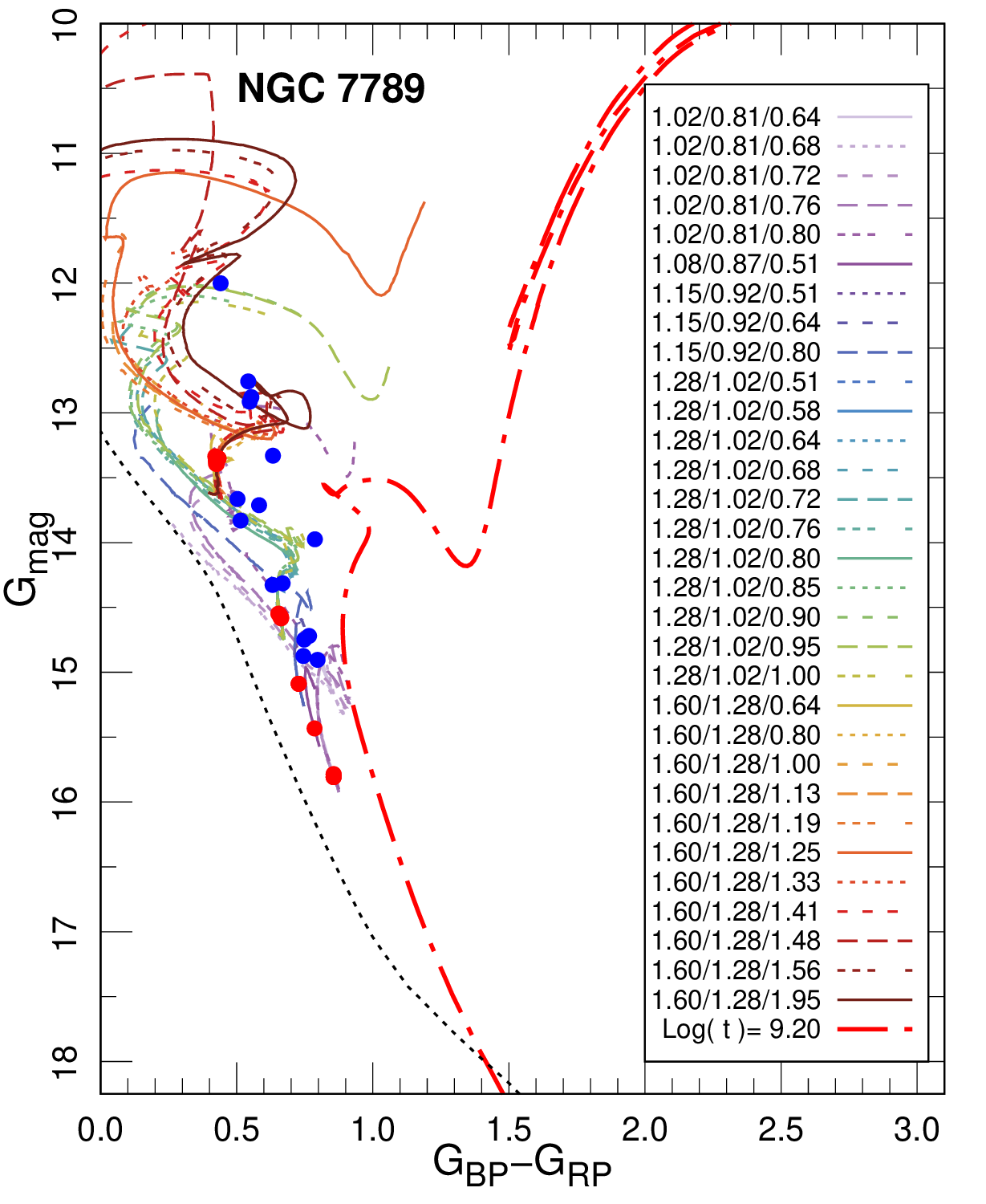}
\caption{}
\label{fig:NGC7789_b}
\end{subfigure}
\caption{Same as Fig.~\ref{fig:six_gaia_hrd1}.}
\label{fig:six_gaia_hrd3}
\end{figure*}

\end{appendix}
\end{document}